\documentclass[preprint,12pt]{elsarticle}

\newtheorem{les}{Lesson}
\usepackage{amssymb}
\usepackage{multirow}
\usepackage{hyperref} 
\usepackage{color}
\usepackage{amsmath}
\usepackage{makecell}
\usepackage{bm}
\usepackage{longtable}
\usepackage{caption}
\newcolumntype{P}[1]{>{\raggedright\arraybackslash\footnotesize}m{#1}}
\newcolumntype{A}[1]{>{\centering\arraybackslash\footnotesize}m{#1}}
\journal{Journal of Network and Computer Applications}

\begin{document}

\begin{frontmatter}

\title{A Survey of Machine Learning-based Physical-Layer Authentication in Wireless Communications}

\author[label1]{Rui Meng}
\author[label1]{Bingxuan Xu}
\author[label1,label2]{Xiaodong Xu\corref{cor1}}
\author[label1]{Mengying Sun}
\author[label1]{Bizhu Wang}
\author[label1]{Shujun Han}
\author[label3]{Suyu Lv}
\author[label1,label2]{Ping Zhang}
\cortext[cor1]{Corresponding author. *Email: xuxiaodong@bupt.edu.cn}

\address[label1]{{State Key Laboratory of Networking and Switching Technology, Beijing University of Posts and Telecommunications}, 
{Beijing}, 
{100876}, 
{China}} 

\address[label2]{{Department of Broadband Communication, Peng Cheng Laboratory}, 
{Shenzhen}, 
{518066}, 
{Guangdong}, 
{China}} 

\address[label3]{{School of Information Science and Technology, Beijing University of Technology}, 
{Beijing}, 
{100124}, 
{China}} 


\begin{abstract}
To ensure secure and reliable communication in wireless systems, authenticating the identities of numerous nodes is imperative. Traditional cryptography-based authentication methods suffer from issues such as low compatibility, reliability, and high complexity. Physical-Layer Authentication (PLA) is emerging as a promising complement due to its exploitation of unique properties in wireless environments. Recently, Machine Learning (ML)-based PLA has gained attention for its intelligence, adaptability, universality, and scalability compared to non-ML approaches. However, a comprehensive overview of state-of-the-art ML-based PLA and its foundational aspects is lacking. This paper presents a comprehensive survey of characteristics and technologies that can be used in the ML-based PLA. We categorize existing ML-based PLA schemes into two main types: multi-device identification and attack detection schemes. In deep learning-based multi-device identification schemes, Deep Neural Networks are employed to train models, avoiding complex processing and expert feature transformation. Deep learning-based multi-device identification schemes are further subdivided, with schemes based on Convolutional Neural Networks being extensively researched. In ML-based attack detection schemes, receivers utilize intelligent ML techniques to set detection thresholds automatically, eliminating the need for manual calculation or knowledge of channel models. ML-based attack detection schemes are categorized into three sub-types: Supervised Learning, Unsupervised Learning, and Reinforcement Learning. Additionally, we summarize open-source datasets used for PLA, encompassing Radio Frequency fingerprints and channel fingerprints. Finally, this paper outlines future research directions to guide researchers in related fields.
\end{abstract}


\end{frontmatter}

\section{Introduction}
\label{sec1}

\subsection{Background}
\label{subsec1}
With the vigorous development of information technology promoted by academia and industry, wireless communication techniques have been widely applied in numerous fields, such as aviation navigation, radio and television, transportation, meteorology, fire prevention, flood control, as well as mobile communications \cite{wang2023road}. According to forecasts, by the year 2025, it is estimated that there will be 7.49 billion mobile users worldwide\footnote{https://www.statista.com/statistics/218984/number-of-global-mobile-users-since-2010/}. However, the misuse of wireless devices for illicit cybercriminal activities has been increasing. This can be attributed to the open and broadcast nature of wireless media, which makes it susceptible to various types of attacks \cite{jin2021introduction}. For instance, malicious users exploit vulnerabilities in wireless networks to eavesdrop on transmitted data and obtain sensitive information such as personal data or trade secrets \cite{moon2019proactive}. They may also deceive unsuspecting users by impersonating legitimate devices, tricking them into sharing sensitive information, or facilitating malicious operations \cite{li2020area}. Additionally, attackers can launch Jamming attacks that disrupt the communication between devices, leading to interruptions, data loss, or degraded communication quality \cite{tan2023detection}. Furthermore, Sybil attackers threaten the reputation and security of wireless networks or systems. These attackers create multiple false identities to manipulate network decision-making processes, monopolize resources, or interfere with the normal functioning of other users \cite{chen2020automated}. The above security threats have caused security threats to many application scenarios, and may even bring serious economic losses. For example, in vehicles ad hoc networks, the dependence on infrastructure, computing, dynamic characteristics and control technology makes its security threats increase \cite{junejo2020privacy,junejo2021lightweight,memon2015secure}. For another example, the security threats of 6G come from the complexity of network architecture, the diversity of access devices, the surge of data traffic and new security threats\cite{ahmed2024secure}.
Therefore, it is crucial for individuals and organizations to be aware of these risks and take appropriate measures to identify wireless devices and guarantee the wireless security.
\subsection{Cryptography-based Upper-Layers Authentication Mechanisms}
Currently, authentication mechanisms in wireless communications are achieved through traditional cryptography-based algorithms at the upper-layers \cite{3gpp2020security}. However, these methods are not applicable for emerging application scenarios, such as the Internet of Things (IoT), the sixth-generation (6G) wireless networks, Industrial Internet of Things (IIoT), and smart cities for the following limitations.
\begin{itemize}
    \item The cryptography-based authentication techniques are based on computational theories (such as algebraic geometry and discrete mathematics) and are realized with one basic assumption that attackers have limited computational capability \cite{xie2022multiple}. However, this assumption has been gradually broken due to the rapid advancements in cryptanalysis algorithms and computational power \cite{xie2022security}. If the root key is leaked, various attacks can compromise the identification system. For example, in the Internet of Vehicles (IoV), malicious nodes can employ Sybil attacks to transmit fake messages, such as incorrect route directions, disturbing networks and posing potential risks to passengers’ lives \cite{sharma2019survey}.
    \item Most cryptography-based approaches are vulnerable to replay attacks, where adversaries can recover the physical-layer bit stream and directly deliver the recovered signal to the legal receiver without modifying the upper-layers signaling or cracking the cryptographic algorithms \cite{xie2022security}. For instance, an attacker may attempt to record transmitted signals from a legitimate transmitter earlier and subsequently replay the recorded signals to pass authentication. This can lead to the legitimate receiver failing to authenticate and disrupt normal communication \cite{oligeri2022past}.
    \item The cryptography-based algorithms necessitate the generation, distribution, and updating of keys, thereby increasing transmission latency \cite{fang2021lightweight}. Hence, they are not suitable for numerous latency-sensitive scenarios \cite{fang2018learning}. For example, health management within intelligent medicine requires patients’ self-management, emphasizing real-time self-monitoring, prompt feedback of health data, and timely medical intervention \cite{tian2019smart}. Additionally, real-time multivariable statistical system monitoring methods are extensively employed in chemical engineering, automobile production, agricultural monitoring, and other industrial sectors \cite{yin2019real}. Failure to guarantee real-time performance may result in significant economic losses and security threats.
    \item The cryptography-based identification methods introduce high communication overhead and complexity, particularly undesired for devices with limited computational and store resources, such as massive machine-type communications and Unmanned Aerial Vehicles (UAV) that are inherently power-limited and processing-restricted \cite{xia2021multiple}. Moreover, due to diminishing compatibility as nodes increase, these approaches struggle to support the ultimate goal of IoT, real-time interaction between things, machines, and people \cite{fang2018learning}. Additionally, with 6G anticipated to support space-air-ground-sea integrated networks encompassing various terminals, divergent encryption and decryption methods between different network protocols pose challenges for devices in achieving swift handovers without service interruption \cite{xie2022security}.  
\end{itemize}

Consequently, more robust and secure identity authentication approaches are required to effectively address the aforementioned limitations of the upper-layers security mechanisms, thus ensuring the wireless security.

\subsection{Physical-Layer Authentication (PLA)}
As a complement of traditional security mechanisms, Physical-Layer Authentication (PLA) has recently been considered a powerful approach for verifying the identity of radio devices due to the below superiorities.
\begin{itemize}
    \item PLA is achieved based on physical-layer features, mainly including radio frequency (RF) fingerprints and channel fingerprints. Such physical-layer attributes are exploited from the communication links, devices, and location-related attributes, and it is challenging for adversaries to extract, imitate, and forge them \cite{xie2022multiple}. In other words, they can provide unique identification signatures and endogenous security for legal devices \cite{li2022blind}.
    \item PLA is a lightweight approach that circumvents many upper-layer signaling processes \cite{wang2016physical}. In addition, since the access point has acquired the Channel State Information (CSI) of all legitimated users during the channel estimation phase, computational overhead is further reduced \cite{xie2021physical}. As a result, radio terminals with finite computing resources can perform effectively \cite{forssell2021worst}.
    \item PLA is highly compatible in heterogeneous coexistence environments \cite{xie2021physical2}. Incompatible devices may not be able to decode each other’s upper-layer signaling, but they should be able to decode physical-layer bit-streams \cite{xie2022multiple}.
\end{itemize}

In earlier literature, PLA-based attack detection is accomplished by formulating a statistical hypothesis test, where the received signal is deemed illegitimate if the difference between the corresponding fingerprint and the reference fingerprint exceeds the threshold; otherwise, it is considered legitimate \cite{xiao2008using}. However, owing to the dynamic and random fluctuations of electromagnetic environments, the impact of noise, and the attackers’ concealment, it is becoming increasingly challenging for non-ML-based PLA methods to determine the theoretical optimal threshold \cite{xiao2016phy}. 

More recently, Machine Learning (ML)-based PLA methods have attracted increased interest. Compared to non-ML-based PLA, ML-based PLA has the following advantages.
\begin{itemize}
    \item 	ML-based PLA is a data-driven method overcoming the challenges in modeling the uncertainty and unknown dynamics of wireless links. For example, for the industrial environments containing machine areas, mobile robot, inspection machine, assembly work cells, and stacking storage area, describing the mathematical expression of the estimated fingerprints of industrial terminals and determining the optimal threshold is not feasible. In this case, we can resort to ML to learn the distribution characteristics and design appropriate algorithms to realize authentication \cite{fang2019machine}.
    \item 	ML-based PLA can realize adaptive threshold authentication. For example, for IoV or UAV scenarios where the channel environments are constantly varying dynamically, the threshold is not always optimal. To address this issue, the receiver can utilize ML algorithms to learn the time-varying physical-layer attributes and realize adaptive online authentication \cite{fang2018learning}.
\item	ML-based PLA is a highly-universal approach without requiring much prior information. For example, with the help of ML techniques, RF fingerprints can be extracted for multi-device identification without the prior-information-dependent expert feature transformation, such as Short Time Fourier Transform (STFT), wavelet transform, and constellation diagram \cite{riyaz2018deep}. For another example, ML-based attack detection can be realized without knowing the prior information of attackers, such as the position and attack frequency \cite{xia2021multiple,shen2021radio23}.
\item	ML-based PLA has higher scalability. Through Transfer Learning (TL) methods, the receiver can quickly identify the test signals of different equipment types in unknown radio environments with only a few training samples on the basis of a pre-training authentication model. In addition, ML-based PLA is an end-to-end authentication process with higher flexibility \cite{yang2021specific, xiang2022review}.
\item	ML-based PLA has the potential to identify large-scale and even ultra-large-scale equipment. ML techniques, especially Deep Learning (DL) methods, are expert in learning high-dimensional features and classifying a large number of samples \cite{chen2019deep,jian2020deep}. In contrast, traditional non-DL approaches, such as feature engineering, can only identify about 100 devices, restricting the development of the Internet of Everything (IoE) \cite{vo2016fingerprinting}.

\end{itemize}

According to the different types of authentication tasks, we categorize the existing ML-based PLA schemes into two categories: multi-device identification and attack detection. 
\begin{itemize}
    \item \emph{Multi-Device Identification:} Most of the state-of-the-art ML-based multi-device identification methods exploit DL techniques to extract the inherent and distinguishable characteristics of RF fingerprints. RF Fingerprint refers to the differences in signal characteristics caused by factors such as device hardware, antennas, and manufacturing processes in wireless communications. These characteristics are unique among devices, analogous to fingerprints in biometrics. Such dissimilarities make the radiation sources of the same model and batch have an inherent property that is different from other individuals \cite{chen2019deep}. Compared the traditional approaches, RF Fingerprint-based methods have the following advantages: no additional hardware required, high uniqueness, good real-time performance, and location tracking.
    DL-based methods can realize intelligent end-to-end identification, while the non-DL-based multi-device identification usually requires much prior information and expert feature transformation to manually set parameters.

\item 	\emph{Attack Detection:} The ML-based attack detection usually considers the conventional “Alice-Bob-Eve” adversarial model and designs how to defend against spoofing attacks or replay attacks. With the help of ML techniques, the detection threshold can be determined automatically without knowing the channel parameters \cite{xiao2017game} or attackers’ information \cite{xia2021multiple}. In contrast, the non-ML-based methods require setting the threshold manually, and it is challenging for the threshold to be adapted to dynamic channel environments.
\end{itemize}

Mukherjee et al. \cite{mukherjee2014principles} provide a survey of Physical Layer Security (PLS) in multiuser wireless networks, and the associated problem of PLA is also briefly discussed. To address the challenges in low reliability of authentication, Wang et al. \cite{wang2016physical} present several promising research areas and provide possible approaches of invoking PLA to reduce the latency. Liu et al. \cite{liu2016physical31} summarize the fundamental theories of PLA, including confidentiality and authentication. Bai et al. \cite{bai2020physical} review the concepts, key techniques as well as future research trends of PLA. Xie et al. \cite{xie2020survey} give a literature survey on passive PLA and active PLA. The active PLA schemes modify the source message on purpose to provide additional identification characteristics, while the passive PLA schemes do not. Angueira et al. \cite{angueira2022survey} present a survey on PLA techniques for ensuring the security in industry, including vulnerabilities, possible attacks, and PLA for factory automation. Xu et al. \cite{xu2015device} provide a tutorial overview of RF fingerprints, including the taxonomy of RF fingerprints, authentication algorithms, and open research problems of fingerprint extraction. Fang et al. \cite{fang2019machine} envision ML-based PLA methods and provide intelligent authentication with a higher security level. The authors of \cite{fan2019rfid,soltani2020more,lee2021deep} develop DL-based PLA schemes for indoor environments with multipath effects, WiFi scenarios, and near field communication (NFC). Jagannath et al. \cite{jagannath2022comprehensive} present a tutorial of DL-based RFF techniques and provide a roadmap of potential research approaches in an illustrative way. Liu et al. \cite{liu2021machine} summarize ML-based identity authentication technologies for IoT devices from the viewpoint of passive surveillance agents and discuss various enabling techniques to secure the IoT. We provide a list of representative overview/survey/tutorial papers on PLA in Tab. \ref{tab1}.

\begin{longtable}{|A{0.6cm}|A{2.2cm}|P{9.5cm}|}  
\caption{\footnotesize{List of Representative Overview/Survey/Tutorial Papers on PLA}} 
\label{tab1}\\
\hline
\textbf{Ref.} & \textbf{Publication Year/Type} & \textbf{Major Contributions} \\
\hline
\endfirsthead

\hline
\textbf{Ref.} & \textbf{Publication Year/Type} & \textbf{Major Contributions} \\
\hline
\endhead

\hline
\endfoot

\hline
\endlastfoot

\cite{xie2022security} & 2022/Overview & Overview different PLS mechanisms, explain the relationship among them and their characteristics, and further introduce several promising approaches to ensure the security. \\
\hline
\cite{wang2016physical} & 2016/Overview & Review PLA techniques, analyze their limitations, provide three promising research areas in dealing with these issues, and further discuss feasible approaches of invoking PLA to reduce the latency. \\
\hline
\cite{fang2019machine} & 2019/Overview & Envision novel PLA approaches based on ML and further introduce different ML paradigms for intelligent and continuous attack detection. \\
\hline
\cite{mukherjee2014principles} & 2014/Survey & Provide a comprehensive survey on PLS based on information-theoretic principles and briefly discuss PLA approaches based on hypothesis testing. \\
\hline
\cite{liu2016physical31} & 2017/Survey & Investigate the fundamental theories of PLS technologies, discuss various PLS techniques and corresponding challenges, and further suggest numerous solutions. \\
\hline
\cite{bai2020physical} & 2020/Survey & Introduce the background, fundamentals, and attack models of PLA, and classify PLA methods into three typical architectures: channel information-based, RF feature-based, and identity watermarks-based. Potential research trends of PLA in multiuser communications are also discussed. \\
\hline
\cite{xie2020survey} & 2021/Survey & Present a comprehensive survey on existing PLA schemes and categorize them into two categories: passive and active schemes. The related works are reviewed in detail. \\
\hline
\cite{angueira2022survey} & 2022/Survey & Give a literature survey on security aspects of industrial wireless communications from industry, academia, and standardization bodies. PLA techniques to defend against spoofing attacks are also reviewed. \\
\hline
\cite{xu2015device} & 2016/Tutorial & Provide a tutorial overview of RFF for enhancing the security of radio networks, including the taxonomy of RF fingerprints and several RFF algorithms. \\
\hline
\cite{fan2019rfid} & 2019/Overview & Review representative literature related to RF fingerprints and research difficulties of multipath effects in indoor radio environments, and further introduce an advanced identification framework based on DL. \\
\hline
\cite{soltani2020more} & 2020/Overview & Review data augmentation approaches that attempt to overcome the drop in RFF accuracy when the channel is dynamically varying between training and testing sets, and further provide two data augmentation methods for enhancing the recognition accuracy. \\
\hline
\cite{lee2021deep} & 2021/Overview & Discuss the feasibility of RF fingerprints used for recognizing NFC tags, implement a hardware testbed for extracting RF features, utilize DL algorithms for experiment, and give key technical challenges. \\
\hline
\cite{jagannath2022comprehensive} & 2022/Tutorial & Provide an elaborated tutorial of traditional and DL-based RFF approaches over the past two decades, including modulation recognition, protocol classification, and emitter identification. \\
\hline
\cite{liu2021machine} & 2022/Survey & Give a survey on the existing techniques on the detection and identification of IoT devices from the perspective of ML, and provide challenges and future research directions for rogue device detection. \\
\hline
\end{longtable}

\subsection{Contributions}

Although numerous researchers focus on ML-based PLA and harness its potential to bolster the identity security of wireless devices, it is astonishing to discover that a comprehensive overview of the state-of-the-art ML-based PLA and its core foundations remains elusive. Consequently, the primary impetus behind this paper is to offer a detailed survey of the characteristics and technologies that can be leveraged within the realm of ML-based PLA. Additionally, the applications of ML-based PLA approaches to various emerging radio communications have recently been proved. Therefore, it is prudent to review the latest cutting-edge ML-based PLA methodologies, which can unveil novel research avenues and directions for researchers in affiliated domains. In this paper, we propose a comprehensive taxonomy for ML-based PLA schemes. The contributions are summarized as follows.

\begin{enumerate}
    \item Initially, we categorize the fingerprints utilized for PLA into two distinct groups: RF fingerprints and channel fingerprints, described as follows.
    \begin{itemize}
        \item RF Fingerprints: These include phenomena such as Carrier Frequency Offset (CFO), In-phase/Quadrature (I/Q) imbalance, and phase noise, which mirror the hardware discrepancies among different devices. Even devices of the same model and batch exhibit unique RF fingerprints.
        
        \item Channel Fingerprints: Encompassing parameters like Received Signal Strength (RSS) and Channel State Information (CSI), these indicators reflect the channel characteristics between transmitters and receivers. The dynamic, time-varying, and richly scattering channel environments furnish distinctive identifying traits for transmitters, known as channel fingerprints.
    \end{itemize}

    \item Subsequently, we classify ML-based PLA schemes into two primary categories: multi-device identification and attack detection.
    \begin{itemize}
        \item PLA for multi-device identification: We compare the non-DL-based and DL-based multi-device identification methods to present the potential and superiority of DL techniques in identification, including not relying on expert feature transformation, end-to-end identification, better scalability, and identification for large-scale and ultra-large-scale devices. We divide the DL techniques for multi-device identification into the following sub-categories: Fully-Connected Neural Networks (FCNN), Convolutional Neural Networks (CNN), Recurrent Neural Networks (RNN), Attention mechanism, data augmentation, Complex-Valued Neural Networks (CVNN), Generative Adversarial Networks (GAN), and Autoencoders (AE). We further provide the architecture of the above-mentioned models and how to extract useful and valuable characteristics of fingerprints, especially raw I/Q fingerprints. Among the DL techniques, CNN is the most widely used model for identification, which are divided into five sub-categories: LeNet-like, AlexNet-like, VGG-like, GoogLeNet-like, and RseNet-like models.

        \item PLA for attack detection: We compare the non-ML-based and ML-based attack detection approaches to present the advantages and advancement of ML technologies in attack detection, including intelligent determination of the optimal threshold and even threshold-free, less dependence on prior information of channel conditions and transmitters, exploitation of multi-fingerprints, and continuous protection. We divide the ML algorithms for attack detection into three sub-categories: Supervised Learning (SL), Unsupervised Learning (UL), and Reinforcement Learning (RL) algorithms. The SL-based methods require the fingerprints and corresponding labels to train the detection system with low false alarm rate and miss detection rate. In contrast, the UL-based approaches require no training fingerprints of attackers, which are more practical in actual wireless communication scenarios. Compared with the SL-based and UL-based methods, the RL-based schemes require no accurate inputs or outputs as well as precise parameter updates. The RL-based detection systems are usually modeled as the game between the legitimate receiver and attackers.

    \end{itemize}

    \item Acknowledging the paramount importance of data in demonstrating the efficacy of ML algorithms, we also summarize open-source datasets of fingerprints to serve as a reference for researchers in related fields.
    \begin{itemize}
        \item The RF fingerprints are outlined based on the number of transmitters, type of receiver and transmitters, waveform, and frequency.
        \item Conversely, the channel fingerprints are summarized according to the provider and channel environments.
    \end{itemize}
    
    \item In addition, we summarize the challenges of existing ML-based PLA schemes and point out the future research direction, including theory, method and practical application.

\end{enumerate}

\subsection{Organization}

\begin{figure*}[h]
\centering
\includegraphics[width=0.95\textwidth]{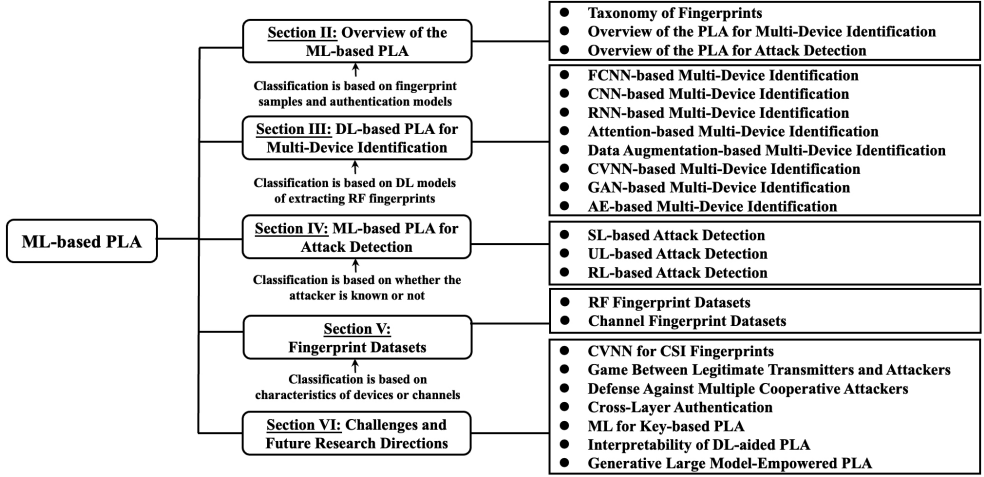}
\caption{Organization of the paper.}
\label{organization of the paper}
\end{figure*}

As illustrated in Fig. \ref{organization of the paper}, the rest of this paper is organized as follows. In Section \ref{Section II}, we provide the taxonomy of fingerprints as well as the comparison of non-ML-based and ML-based PLA. The DL-based PLA for multi-device identification and ML-based PLA for attack detection are comprehensively presented in Section \ref{Section III} and Section \ref{Section IV}, respectively. In Section \ref{Section V}, we summarize open-source datasets of fingerprints. Section \ref{Section VI} and Section \ref{Section VII} respectively show future research directions and conclusions. The acronyms used in this paper are listed in Tab. \ref{tab2}.

\begin{longtable}{|A{2.4cm}|A{3.5cm}|A{2.4cm}|A{3.5cm}|}
\caption{\footnotesize{List of Acronyms Used in the Paper}} 
\label{tab2}\\
\hline
\textbf{Abbreviations} & \textbf{Full Name} & \textbf{Abbreviations} &\textbf{Full Name}
\\
\hline
\endfirsthead
\hline
\textbf{Abbreviations} & \textbf{Full Name} & \textbf{Abbreviations} &\textbf{Full Name}\\
\hline
\endhead

\hline
\endfoot

\hline
\endlastfoot
6G            & The sixth-generation                       & IoT           & Internet of Things                         \\ \hline
ADS-B         & Automatic-dependent surveillance-broadcast & IoV           & Internet of Vehicles                       \\ \hline
AE            & Autoencoder                                & KNN           & K-Nearest Neighbor                         \\ \hline
AoA           & Angle of arrival                           & LDA           & Linear Discriminant Analysis               \\ \hline
BN            & Batch Normalization                        & LFDA          & Linear Fisher Discriminant Analysis        \\ \hline
CAA           & Chaotic Antenna Array                      & LLRT          & Logarithmic likelihood ratio test          \\ \hline
CFO           & Carrier Frequency Offset                   & LSTM          & Long Short-Term Memory                     \\ \hline
CFR           & Channel Frequency Response                 & LTE           & Long-Term Evolution                        \\ \hline
CIR           & Channel Impulse Response                   & MIMO          & Multiple Input Multiple Output             \\ \hline
CLRT          & Classical Likelihood Ratio Test            & ML            & Machine Learning                           \\ \hline
CNN           & Convolutional Neural Network               & MSCNN         & Multi-Scale Convolutional Neural Network   \\ \hline
CSI           & Channel State Information                  & NFC           & Near field communication                   \\ \hline
CVNN          & Complex-Valued Neural Network              & OFDM          & Orthogonal Frequency Division Multiplexing \\ \hline
CWD           & Choi-Williams Distribution                 & PLA           & Physical-Layer Authentication              \\ \hline
DAC           & Digital-to-Analog Converter                & PLS           & Physical-Layer Security                    \\ \hline
DL            & Deep Learning                              & PSD           & Power spectral density                     \\ \hline
DNN           & Deep Neural Network                        & PUWS          & Physically unclonable wireless system      \\ \hline
DRL           & Deep Reinforcement Learning                & ReLU          & Rectified Linear Unit                      \\ \hline
DT            & Decision Tree                              & RF            & Radio Frequency                            \\ \hline
EI            & Edge Intelligence                          & RFF           & Radio Frequency Fingerprinting             \\ \hline
ELM           & Extreme Learning Machine                   & RL            & Reinforcement Learning                     \\ \hline
FCNN          & Fully-Connected Neural Network             & RNN           & Recurrent Neural Network                   \\ \hline
FFT           & Fast Fourier Transform                     & RSS           & Received Signal Strength                   \\ \hline
FHSS          & Frequency hopping spread spectrum          & RSSI          & Received Signal Strength Indication        \\ \hline
FL            & Federated Learning                         & RVNN          & Real-Valued Neural Network                 \\ \hline
GAN           & Generative Adversarial Network             & SEI           & Specific Emitter Identification            \\ \hline
GCN           & Graph Neural Network                       & SL            & Supervised Learning                        \\ \hline
GLRT          & Generalized likelihood ratio test          & SNR           & Signal-Noise Ratio                         \\ \hline
GMM           & Gaussian Mixture Model                     & STFT          & Short Time Fourier Transform               \\ \hline
GP            & Gaussian Process                           & SVM           & Support Vector Machine                     \\ \hline
GPC           & Gaussian Process Classification            & TL            & Transfer Learning                          \\ \hline
GPR           & Gaussian Process Regression                & UAV           & Unmanned Aerial Vehicle                    \\ \hline
GRU           & Gated Recurrent Unit                       & UL            & Unsupervised Learning                      \\ \hline
HHT           & Hilbert-Huang Transform                    & UWSN          & Underwater acoustic sensor network         \\ \hline
I/Q           & In-phase/Quadrature                        & VAE           & Variational Autoencoder                    \\ \hline
IAT           & Inter arrival time                         & VANET         & Vehicular Ad Hoc Network                   \\ \hline
IIoT          & Industrial Internet of Things              & WVD           & Wegener-Ville Distribution                 \\ \hline
\end{longtable}

\section{Overview of the ML-based PLA}
\label{Section II}
\begin{figure*}[h]
\centering
\includegraphics[width=0.95\textwidth]{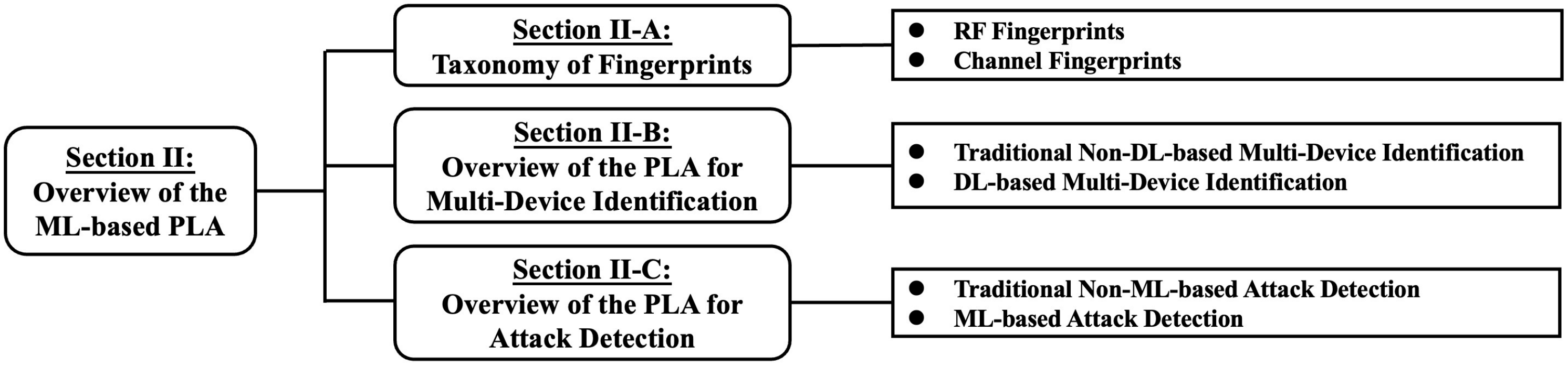}
\caption{Organization of Section II.}
\label{Organization of Section II}
\end{figure*}

In this section, we introduce the overview of the ML-based PLA, including the taxonomy of fingerprints, overview of the non-DL-based and DL-based multi-device identification, and overview of the non-ML-based and ML-based attack detection. The organization of this section is illustrated in Fig. \ref{Organization of Section II}.

\subsection{Taxonomy of Fingerprints}
We divide the fingerprints used for PLA into two categories: RF fingerprints and channel fingerprints. The RF fingerprints are extracted based on the hardware differences of transmitters, including Digital-to-Analog Converter (DAC), I/Q modulator, filter, and power amplifier. Such dissimilarities make the radiation sources of the same model and batch have an inherent property that is different from other individuals, and we call it RF fingerprint \cite{oligeri2022past,hamdaoui2022deep}. In contrast, channel fingerprints are extracted based on the wireless channel environments and reflect the channel characteristics between the transmitter and receiver, including path loss, shadowing effects, and small-scale fading. The detailed descriptions of RF fingerprints and channel fingerprints are as follows, and we provide a comparison between them in Tab. \ref{Taxonomy of Fingerprints Used for PLA} for more clarity.

\begin{table*}[h]
\caption{Taxonomy of Fingerprints Used for PLA}
\label{Taxonomy of Fingerprints Used for PLA}
\center
\begin{tabular}{|p{1cm}|A{2cm}|A{4cm}|A{4cm}|}
\hline
\multicolumn{2}{|c|}{\textbf{Types of Fingerprints}}                                              & \textbf{Physical-Layer Attributes}& \textbf{Application Scenarios}\\ \hline

\multicolumn{1}{|c|}
{\multirow{2}{*}[-5ex]{ \begin{tabular}[c]{@{}A{2cm}@{}} RF fingerprints\end{tabular}}} & Transient fingerprints     & Wavelet coefficient \cite{choe1995novel,toonstra1996radio}, transient amplitude \cite{hall2004enhancing,tekbacs2004improvement}, and PSD \cite{tekbacs2004improvement,suski2008using}   & \multirow{2}{*}{\begin{tabular}[c]{@{}A{4cm}@{}} OFDM systems \cite{hou2014physical}, \\ Bluetooth networks \cite{faria2006detecting}, \\ZigBee network \cite{dubendorfer2012rf}, WiMax system \cite{reising2012wimax}, Ad-Hoc network \cite{ureten2007wireless}, IoT \cite{chatterjee2018rf,zhang2022adaptive}, UAVs \cite{reus2021classifying}, etc.\end{tabular}}                                             \\ \cline{2-3}
\multicolumn{1}{|c|}{}       & Steady-state fingerprints  & I/Q imbalance \cite{hao2014performance,hao2014relay,sankhe2019oracle}, Carrier Frequency Offset (CFO)\cite{hou2012physical,hou2014physical,zeng2018physical}, clock skew \cite{kohno2005remote,jana2008fast,cristea2013fingerprinting}, phase noise \cite{pitarokoilis2015ml,zhao2017robust}, and imperfect power amplifier \cite{dolatshahi2010identification,polak2011identifying,polak2015identification}&               
\\ \hline
\multicolumn{1}{|c|}{\multirow{2}{*}[-5ex]{{\begin{tabular}[c]{@{}A{2cm}@{}} Channel fingerprints\end{tabular}}}} & Statistical fingerprints   & Received Signal Strength (RSS) \cite{varshavsky2007amigo,kalamandeen2010ensemble,zhong2004privacy}, Received Signal Strength Indication (RSSI) \cite{fang2018learning,demirbas2006rssi},  angle of arrival (AoA) \cite{xu2018independence,abdelaziz2019enhanced}, and PSD \cite{corbett2007passive,kennedy2008radio,williams2010rf}                          & \multirow{2}{*}[3ex]{\begin{tabular}[c]{@{}A{4cm}@{}}MIMO systems \cite{xiao2008mimo},  dual-hop wireless networks \cite{zhang2019end}, OFDM systems \cite{he2010ream}, CDMA systems \cite{he2009epson}, wireless sensor networks \cite{zhong2004privacy}, industrial cyber-physical systems \cite{pan2019threshold}, Mobile Edge Computing(MEC) \cite{liao2019security}, etc.\end{tabular}}\\ \cline{2-3}
\multicolumn{1}{|c|}{}                                      & Instantaneous fingerprints & Channel State Information (CSI) \cite{wang2015privacy,xu2018phy}, Channel Impulse Response (CIR) \cite{tugnait2010channel,liu2011robust}, and Channel Frequency Response
(CFR)\cite{xiao2007fingerprints,xiao2008using} & \\ \hline
\end{tabular}
\end{table*}

\subsubsection{RF Fingerprints}
We divide RF fingerprints into two sub-categories: transient fingerprints and steady-state fingerprints. The former sub-categories reflect the response of the transmitter elements when they are subjected to transient impulses such as startup and shutdown, and contain rich nonlinear and non-stationary characteristics. The latter one reflects the features extracted in the signal modulation phase. Transient fingerprints contain wavelet coefficient \cite{choe1995novel,toonstra1995transient}, transient amplitude \cite{hall2004enhancing,barbeau2006detection,tekbacs2004improvement}, and transient power spectral density (PSD) \cite{tekbacs2004improvement,suski2008using}, while steady-state fingerprints mainly contain I/Q imbalance \cite{hao2014performance,sankhe2019oracle}, CFO \cite{hou2012physical,hou2014physical,zeng2018physical}, clock skew \cite{kohno2005remote,jana2008fast,cristea2013fingerprinting}, phase noise \cite{pitarokoilis2015ml,zhao2017robust}, and imperfect power amplifier \cite{dolatshahi2010identification,polak2011identifying,polak2015identification}. The open-source datasets of RF fingerprints will be introduced in Section \ref{Section V-A} in detail.
\subsubsection{Channel Fingerprints}
We divide channel fingerprints into two sub-categories: statistical fingerprints and instantaneous fingerprints. The former sub-categories indicate the statistical information of wireless links, while the latter one represents the fine-grained features of radio channels. Statistical fingerprints mainly contain RSS \cite{varshavsky2007amigo,kalamandeen2010ensemble,zhong2004privacy}, Received Signal Strength Indication (RSSI) \cite{demirbas2006rssi,fang2018learning}, angle of arrival (AoA) \cite{xu2018independence,abdelaziz2019enhanced}, and PSD \cite{corbett2007passive,kennedy2008radio,williams2010rf}. Instantaneous fingerprints include CSI \cite{wang2015privacy,xu2018phy}, Channel Impulse Response (CIR) \cite{tugnait2010channel,liu2011robust}, and Channel Frequency Response (CFR) \cite{xiao2007fingerprints,xiao2008using}. The open-source datasets of channel fingerprints will be introduced in Section \ref{Section V-B} in detail.
\subsection{Overview of the PLA for Multi-Device Identification}
Most works related to PLA for multi-device identification adopt RF fingerprints as the identity signatures of transmitters. The RF fingerprint-based multi-device identification is also known as radio frequency fingerprinting (RFF) \cite{shen2021radio23,xie2021generalizable} or specific emitter identification (SEI) \cite{wang2022few,yang2021specific}. The object of multi-device identification is to recognize which transmitter in the fingerprint database matches the received signal in the authentication phase, and the identification process is as follows.
\begin{enumerate}[Step 1)]
    \item Collect signals of devices;
\item Pre-Process signals;
\item Further process signals by expert feature transformation;
\item Train the classifier for authentication;
\item Identify unknown signals using the trained classifier.
\end{enumerate}

The most widely used metric is the identification accuracy denoted by (1).
\begin{equation}
AucRate =\frac{1}{N}\sum_{n=1}^{N}\mathbb{I}(\bm{L}_{n}=\bm{Y}_{n})
\end{equation}
where $N$ denotes the number of fingerprints in the authentication phase. $\bm{L}_n$ and $\bm{Y}_n$ denote the real label and predicted label of the $n$th fingerprint, respectively. $\mathbb{I}(\cdot)$ denotes the indicator function, where if $\cdot$ is true, it takes 1; otherwise, it takes 0. $AucRate$ values range from 0 to 1. When all the fingerprint samples are identified correctly, $AucRate$ takes 1.

Here, we compare the traditional non-DL-based and DL-based multi-device identification schemes.
\subsubsection{Traditional Non-DL-based Multi-Device Identification}
"Step 2" includes two parts. The first part refers to pre-processing without prior information, such as normalization, interpolation, and Fast Fourier Transform (FFT), while the second part refers to preprocessing with prior information, such as time synchronization and phase offset compensation. Moreover, the expert feature transformation in “Step 3”, such as STFT \cite{lopez2005digital}, wavelet transform \cite{bertoncini2011wavelet}, Wegener-Ville Distribution (WVD) \cite{lunden2007automatic}, Choi-Williams Distribution (CWD) \cite{lunden2007automatic}, High-order spectrum \cite{tugnait1994detection,chandran1993pattern,zhang2001new}, Hilbert-Huang Transform (HHT) \cite{dragomiretskiy2013variational,zhang2015novel,zhang2016specific}, also require prior information to manually set parameters.
\subsubsection{DL-based Multi-Device Identification}
The DL-based approaches are achieved through Deep Neural Networks (DNN), including FCNN \cite{roy2019detection}, CNN \cite{sankhe2019oracle,riyaz2018deep,aneja2018iot,o2016convolutional,hermawan2020cnn,o2018over,shen2021radio105}, RNN \cite{jafari2018iot,wu2018deep,he2020cooperative,al2021deeplora}, Transformer \cite{shen2021radio114,xu2021transformer}, data augmentation \cite{soltani2020more,al2021deeplora,shen2021radio114,meng2022multiuser,liu2020specific,cekic2021wireless,gul2022fine}, CVNN \cite{agadakos2020chameleons,gopalakrishnan2019robust,wang2021efficient,brown2021charrnets}, Graph Neural Networks (GNN) \cite{li2023novel}, GAN \cite{zhao2018classification,roy2019detection}, and AE \cite{karunaratne2021open,yu2019radio,bassey2020device}. 
For instance, \cite{roy2019detection} proposes a FCNN-based framework in which biases and regularization techniques are employed to mitigate underfitting and overfitting. The Adam optimization algorithm is utilized for gradient descent to update the network parameters. \cite{sankhe2019oracle} introduces a CNN classifier called ORACLE to analyze the differences between I/Q samples from a large pool of bit-similar devices. The ORACLE architecture consists of two convolutional layers and two fully-connected layers. Each complex I/Q sample, serving as the input to the ORACLE model, is represented as a 2-dimensional real value.
\cite{al2021deeplora} evaluates the performance of CNN (1D CNN and 2D CNN) and LSTM using various metrics: ‘Per-slice Training’ accuracy, ‘Per-slice Testing’ accuracy, ‘Train-and-Test-Same-Day’ accuracy, and ‘Train-and-Test-Other-Day’ accuracy. \cite{shen2021radio114} proposes a transformer model for identifying LoRa devices using signals of variable lengths and introduce a multi-packet inference approach to significantly enhance accuracy in low SNR environments. Unlike adding Gaussian noise at the input layer, \cite{meng2022multiuser} proposes an LPNN approach to ensure the security of IIoT access. Given that latent layers exhibit stronger linear characteristics than input layers for CSI fingerprints, the proposed LPNN method offers better interpretability. \cite{wang2021efficient} develops an efficient approach using CVNN and network compression, named SlimCVNN, to achieve both high accuracy and low model complexity. \cite{roy2019detection} employs GANs for recognizing adversarial fingerprints and identifying wireless devices. Their generative module can produce fake fingerprints, while the discriminative module distinguishes real from fake ones. The CNN module enables the classification of legitimate devices. \cite{bassey2020device} proposes an AE-based architecture for intrusion detection and introduce the concept of a device authentication code. The reconstruction error serves as the device’s authentication code, and the Kolmogorov-Smirnov test is used to determine the legitimacy of fingerprints.

For the DL-based schemes, “Step 2” only includes the above-mentioned first part, and “Step 3” is needless. In other words, the DL-based multi-device identification can directly use raw fingerprints to realize end-to-end identity authentication, overcoming the challenges in obtaining prior information and optimizing parameters manually.

\begin{figure*}[h]
\centering
\includegraphics[width=0.95\textwidth]{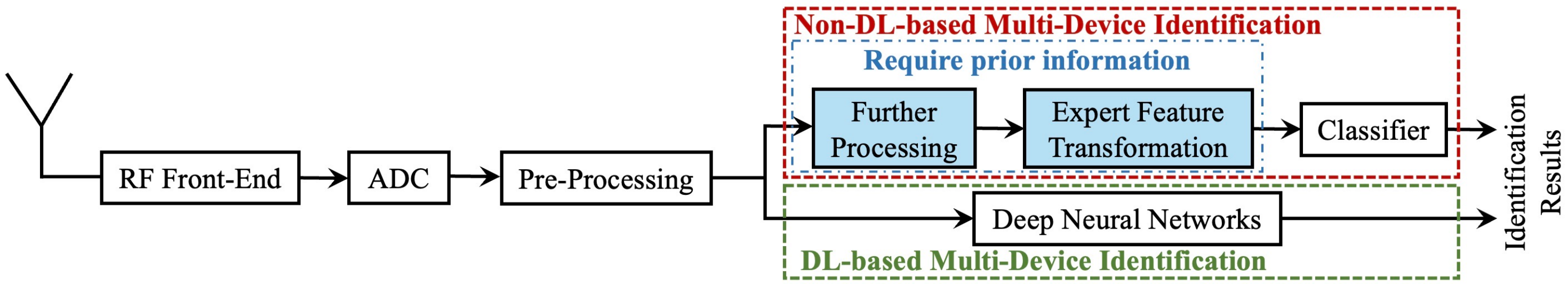}
\caption{Illustration of the non-DL-based and DL-based multi-device identification methods.}
\label{Illustration of the non-DL-based and}
\end{figure*}

We further provide Fig. \ref{Illustration of the non-DL-based and} and Tab. \ref{Comparison of the Non-DL-based and DL-based Multi-Device Identification Methods} to illustrate the comparison of the traditional non-DL-based and DL-based multi-device identification schemes, which are beneficial for readers to understand more clearly. The DL-based methods will be introduced in Section \ref{Section III} in detail.

\newpage
\begin{longtable}{|P{1.8cm}|A{1.8cm}|P{6.2cm}|P{2.1cm}|}
\caption{Comparison of the Non-DL-based and DL-based Multi-Device Identification Methods} \label{Comparison of the Non-DL-based and DL-based Multi-Device Identification Methods} \\
\hline
\textbf{Categories} & \textbf{Sub-Categories} & \textbf{Descriptions} & \textbf{Advantages and Disadvantage } \\ \hline
\endfirsthead 
\hline
\textbf{Categories} & \textbf{Sub-Categories} & \textbf{Descriptions} & \textbf{Advantages and  Disadvantages} \\
\hline
\endhead

\hline
\endfoot

\hline
\endlastfoot

\hline
\multirow{3}{*}[-10ex]{\begin{tabular}[c]{@{}A{2cm}@{}}  Non-DL-based multi-device identification\end{tabular}} & Time-frequency transformation \cite{lopez2005digital,bertoncini2011wavelet,lunden2007automatic}. &Describe the time variation law and frequency distribution of signals, including linear time-frequency transformation (STFT \cite{lopez2005digital}, Wavelet transform \cite{bertoncini2011wavelet}) and quadratic time-frequency transformation (WVD \cite{lunden2007automatic}, CWD \cite{lunden2007automatic}). & \multirow{3}{*}{\begin{tabular}[c]{@{}P{2cm}@{}}Require more prior information and expert feature transformation to manually set parameters, but have better interpretability.\end{tabular}}\\ \cline{2-3}

 & High-order spectrum \cite{tugnait1994detection,chandran1993pattern,zhang2001new}.  & Reserve unintentional modulation information and effectively suppress additive Gaussian noise, including bispectrum, integral poly-spectrum, and rectangular integral bispectrum.& \\ \cline{2-3}& Hilbert spectrum \cite{dragomiretskiy2013variational,zhang2015novel,zhang2016specific}.       & Show the law of time variation and frequency distribution of signals, and interpret the relationship between time and frequency of signals, mainly including HHT. & \\ \hline

\multirow{8}{*}[-10ex]{\begin{tabular}[c]{@{}A{1.8cm}@{}}DL-based multi-device identification\end{tabular}}   & FCNN  \cite{roy2019detection} &The FCNN models are fully connected. With the increase of dataset size, it is more and more difficult to obtain robust fingerprints.        & \multirow{8}{*}[-2ex]{\begin{tabular}[c]{@{}P{2.1cm}@{}}Require no prior information, realize higher accuracy, and achieve end-to-end identification, but require more training datasets and have worse interpretability.\end{tabular}} \\ \cline{2-3}& CNN\cite{riyaz2018deep,sankhe2019oracle},\cite{aneja2018iot,o2016convolutional,hermawan2020cnn}   & The CNN models exploit convolutional layers and pooling layers to extract features of fingerprints, including LeNet-like models \cite{riyaz2018deep,sankhe2019oracle,o2016convolutional}, ResNet-like models \cite{al2020exposing}, \cite{han2023radar}, AlexNet-like models \cite{al2020exposing,sankhe2019no}, VGG-like models\cite{elmaghbub2020leveraging}, etc. &\\ \cline{2-3}& \begin{tabular}[c]{@{}A{2cm}@{}}RNN\cite{jafari2018iot,wu2018deep,he2020cooperative}\end{tabular}& The RNN models can mine time-related features,including Long Short-Term Memory (LSTM)  models \cite{jafari2018iot,wu2018deep} and the combinations of CNN and LSTM models  \cite{al2021deeplora}, \cite{liu2020specific}. & \\ \cline{2-3}& Data augmentation \cite{soltani2020more,al2021deeplora,shen2021radio114},\cite{meng2022multiuser,liu2020specific,cekic2021wireless,gul2022fine}               & The data augmentation methods can address the problem of inconsistent probability distribution between training fingerprints and testing fingerprints, avoid over-fitting issues, and improve the generalization of models, including adding Gaussian noise \cite{al2021deeplora,shen2021radio114}, \cite{meng2022multiuser} and generating new samples \cite{soltani2020more,al2021deeplora},\cite{liu2020specific,cekic2021wireless,karunaratne2021open}. & \\ \cline{2-3}& CVNN\cite{agadakos2020chameleons,gopalakrishnan2019robust,wang2021efficient,brown2021charrnets}. & The CVNN models can directly process complex baseband signals and are composed of complex-valued convolution layers, complex-valued fully-connected layers, complex-valued batch normalization, and complex-valued activation functions.& \\ \cline{2-3}
& GAN \cite{roy2019detection,zhao2018classification}. & The GAN models can realize the classification of multiple trusted transmitters and the identification of rogue devices through generators and discriminators. & \\ \cline{2-3}& AE\cite{karunaratne2021open,yu2019radio,bassey2020device}                            & The AE models can achieve feature extraction and dimension reduction through encoders and decoders. & \\ \hline
\end{longtable}

\subsection{Overview of the PLA for Attack Detection}
Among the works related to PLA for attack detection, some only use RF fingerprints as identity signatures of legitimate devices, some only use channel fingerprints to construct the hypothesis testing, and the rest utilize the combination of channel fingerprints and RF fingerprints to obtain more robust detection performance. Considering that Xie et al. \cite{xie2020survey} have provided a comprehensive survey of different types of fingerprints for attack detection, we focus on enabling ML methods. Hence, we summarize the works on attack detection from the perspective of ML techniques.
The goal of attack detection at the legitimate receiver (Bob) is to identify whether the received signal is legal (from Alice) or not (from Eve), and the detection can be realized using the following hypothesis testing in (2).
\begin{equation}
\left\{\begin{matrix}\mathcal{H}_0\colon H_{diff}=diff\big(H_t-H_{ref}\big)<\theta\\\mathcal{H}_1\colon H_{diff}=diff\big(H_t-H_{ref}\big)\geq\theta\end{matrix}\right.
\end{equation}
where $\mathcal{H}_0$ and $\mathcal{H}_1$ respectively indicate the signal corresponding to the unknown fingerprint $H_t$ comes from Alice and Eve. $H_{ref}$ represents the reference fingerprint, $diff(H_t-H_{ref})$ is the difference between $H_t$ and $H_ref$, and $\theta$ represents the threshold for comparison. $\theta$ is a critical optimization parameter. If the value of $\theta$ is too small, the authentication system will become overly sensitive, leading to the misidentification of some legitimate signals as illegal. On the other hand, if the value of $\theta$ is too large, the system will react too loosely, resulting in the misclassification of some illegal signals as legitimate.

The common metric for evaluating the detection performance includes false alarm rate $P_{fa}$ and miss detection rate $P_{md}$, which can be represented as (3) and (4), respectively.
\begin{equation}
P_{fa}=P\{\hat{y}=0|y=1\}
\end{equation}
\begin{equation}
P_{md}=P\{\hat{y}=1|y=0\}
\end{equation}
where $\widehat{y}$ is the predicted identity, and $y$ is the real identity. $P_{fa}$ denotes the probability of legal fingerprints being falsely alarmed as the spoofing attack, and $P_{md}$ represents the probability of spoofing signals being missed detection. Therefore, $\theta$ in (2) are optimized according to the trade off between $P_{fa}$ and $P_{md}$.

\begin{figure*}[h]
\centering
\includegraphics[width=0.95\textwidth]{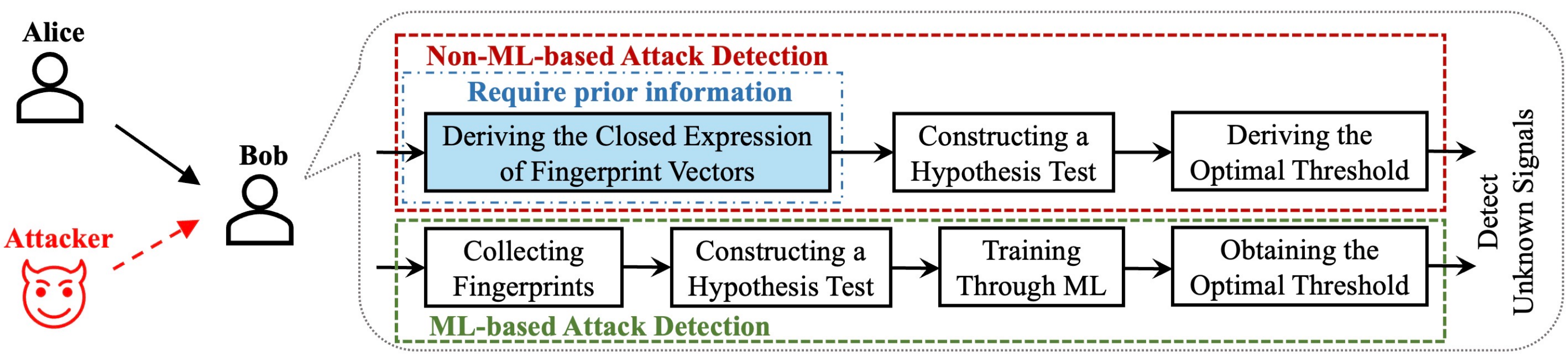}
\caption{Illustration of the non-ML-based and ML-based attack detection methods.}
\label{Illustration of the non-ML-based and}
\end{figure*}

Here, we compare the traditional non-ML-based and ML-based attack detection schemes. We further provide Fig. \ref{Illustration of the non-ML-based and} and Tab. \ref{Comparison of the Non-ML-based and ML-based Attack Detection Methods} to illustrate the comparison of the traditional non-ML-based and ML-based attack detection schemes more clearly.

\begin{longtable}{|A{1.8cm}|A{0.5cm}|P{1.5cm}|P{5.7cm}|P{2cm}|}
\caption{Comparison of the Non-ML-based and ML-based Attack Detection Methods} \label{Comparison of the Non-ML-based and ML-based Attack Detection Methods} \\
\hline
\textbf{Categories} & \multicolumn{2}{|A{2cm}|}{\textbf{Sub-Categories}} & \textbf{Descriptions} & \textbf{Advantages and Disadvantages} \\ \hline
\endfirsthead 
\hline
\textbf{Categories} & \multicolumn{2}{|A{2cm}|}{\textbf{Sub-Categories}} & \textbf{Descriptions} & \textbf{Advantages and Disadvantages} \\
\hline
\endhead

\hline
\endfoot

\hline
\endlastfoot

\hline
\multirow{8}{*}[-15ex]{\begin{tabular}[c]{@{}A{1.8cm}@{}}Non-ML-based attack detection\end{tabular}} & \multicolumn{2}{|P{2cm}|}{Fingerprint matching\cite{varshavsky2007amigo,kalamandeen2010ensemble},\cite{faria2006detecting}} & Defend against both the impersonation attack and Sybil attack by matching the estimated fingerprints with the legitimate one.& \multirow{8}{*}[-12ex]{\begin{tabular}[c]{@{}P{2cm}@{}}Better interpretability, but require prior information of transmitters and channels and the determined threshold is unable to adapt to the dynamic channel environment.\end{tabular}} \\ \cline{2-4}& \multicolumn{2}{|P{2cm}|}{Fingerprint ratio\cite{demirbas2006rssi}} & Using RSSI values directly for Sybi attack detection is unfeasible, and a promising method is to use the RSS ratio.&\\ \cline{2-4}
 & \multicolumn{2}{|P{2cm}|}{\begin{tabular}[c]{@{}P{2cm}@{}}Fingerprint similarity\\\cite{zeng2010non,xiao2008physical}\end{tabular}} & Fingerprint similarity can be calculated based on the correlation between the fingerprints of different transmitters.&\\ \cline{2-4}
  &  \multicolumn{2}{|P{2cm}|}{\begin{tabular}[c]{@{}P{2cm}@{}}Residual algorithm\\\cite{malaney2005securing}\end{tabular}}& The RSS residual measured by the Euclidean distance can be used to recognize threats from a malicious user.&\\ \cline{2-4}
&\multicolumn{2}{|P{2cm}|}{\begin{tabular}[c]{@{}P{2cm}@{}}Two-dimensional quantization algorithm\\\cite{liu2013two,liu2016physical}\end{tabular}} & The CIR fingerprints can be quantized in two dimensions for more reliable spoofing detection.& \\ \cline{2-4}
& \multicolumn{2}{|P{2cm}|}{GLRT\cite{xiao2009channel}}  & GLRT is optimal for frequency-selective Rayleigh channels but computationally cumbersome. & \\ \cline{2-4}
  &  \multicolumn{2}{|P{2cm}|}{CLRT\cite{dolatshahi2010identification}}& CLRT can be used to address the detection issue resulted from GLRT.& \\ \cline{2-4}
 &  \multicolumn{2}{|P{2cm}|}{LLRT\cite{liu2013two}}& LLRT is developed based on the fingerprints estimated by the two-dimensional quantization algorithm.& \\ \hline

 \multirow{17}{*}[20ex]{\begin{tabular}[c]{@{}A{1.8cm}@{}}ML-based attack detection\end{tabular}} & \multirow{9}{*}[8ex]{\begin{tabular}[c]{@{}A{0.5cm}@{}}SL\end{tabular}}                      & \begin{tabular}[c]{@{}P{1.5cm}@{}}Logistic regression\\ \cite{xiao2017phy}                     \end{tabular}& Logistic regression is a linear regression analysis model. & \\ \cline{3-4}
 & & LDA\cite{wang2021safeguarding}  & LDA is a linear learning method for detecting multi-attackers.&  \\ \cline{3-4}
 & & DT\cite{enad2020machine,pan2019threshold} & DT is a tree structure based on conditional probability and can intuitively show the detection results. &  \multirow{17}{*}[-13ex]{\begin{tabular}[c]{@{}P{2cm}@{}}Better attack detection performance without manually determining the threshold, but has worse interpretability.\end{tabular}}\\ \cline{3-4}&  &KNN\cite{senigagliesi2022authentication} & In KNN-based schemes, each fingerprint can be represented by its closest neighboring fingerprints. &\\ \cline{3-4}
  &  &SVM\cite{liu2017authenticating,abdrabou2022adaptive}.& SVM can obtain detection results by maximum-margin hyperplane and use kernel methods to realize nonlinear classification. &\\ \cline{3-4}& & \begin{tabular}[c]{@{}P{2cm}@{}}Ensemble learning\\\cite{pan2019threshold,xie2021weighted}\end{tabular} & Ensemble learning can construct multi-classifiers and combine them to realize more robust attack detection performance.& \\ \cline{3-4}& & ELM\cite{wang2017physical} & ELM is a kind of feedforward neuron network structure without updating weights and can realize better generalization in detection than SVM.&\\ \cline{3-4}
   & & FL\cite{wang2021collaborative}  & FL is a distributed ML technique and can realize collaborative authentication without exchanging their estimated fingerprints. &\\ \cline{3-4}&                          & DL\cite{liao2019deep,chen2021physical}                                      & DL can mine the depth characteristics of fingerprints.   &\\ \cline{2-4} &
   
   {\multirow{2}{*}[-10ex]{UL}}
   & \begin{tabular}[c]{@{}P{2cm}@{}}Clustering\\ \cite{xia2021multiple}\end{tabular} & Clustering can divide the fingerprints into several disjoint clusters to achieve attack detection.&\\ \cline{3-4}& & OCC-SVM\cite{abdrabou2022adaptive} & OCC-SVM can leverage kernel functions to map the original feature space of fingerprints to a higher dimensional space, so as to find an effective boundary to separate malicious fingerprints.&\\ \cline{3-4} &  & \begin{tabular}[c]{@{}P{1.5cm}@{}}Manifold learning\\\cite{xia2022physical} \end{tabular}                      & Manifold learning can recover the low-dimensional manifold structure from the high-dimensional fingerprints and find the corresponding embedded mapping to realize detection.&\\ \cline{3-4} & & GMM\cite{gulati2013gmm} & GMM models the fingerprints as Gaussian distributions. &\\ \cline{3-4}   &   & GP\cite{wang2021channel}& GP mainly includes GP Regression (GPR) and GP Classification (GPC).  &\\ \cline{3-4}& & AE\cite{meng2022physical}                                       & AE can realize dimension reduction and further identify the latent vectors. &\\ \cline{2-4}& {\multirow{2}{*}[-6ex]{RL}}                       & Non-DL-aided RL\cite{gao2020physical,xiao2016phy}  & RL can use dynamic game without knowing the system parameters to obtain the optimal threshold. &\\ \cline{3-4} & & DRL\cite{lu2020reinforcement,wu2023game}                                     & DRL can combine the perception ability of DL with the decision-making ability of RL to find better authentication modes and parameters. &\\ \hline
\end{longtable}

\subsubsection{Traditional Non-ML-based Attack Detection}
The non-ML-based attack detection uses the prior information of transmitters and channels to derive the probability distribution of fingerprints, then constructs hypothesis testing statistics, and finds the optimal threshold. The common methods include fingerprint matching \cite{varshavsky2007amigo,kalamandeen2010ensemble,faria2006detecting}, fingerprint ratio \cite{demirbas2006rssi}, fingerprint similarity \cite{zeng2010non,xiao2008physical}, residual algorithm \cite{malaney2005securing}, two-dimensional quantization algorithm \cite{liu2013two,liu2016physical}, generalized likelihood ratio test (GLRT) \cite{xiao2009channel}, classical likelihood ratio test (CLRT) \cite{dolatshahi2010identification}, and logarithmic likelihood ratio test (LLRT) \cite{liu2013two}.
\subsubsection{ML-based Attack Detection}
The ML-based attack detection exploits intelligent ML algorithms to automatically obtain the threshold without manual operations. Some schemes even achieve threshold-free authentication \cite{pan2019threshold}. We classify the ML techniques into three sub-categories: Supervised Learning (SL), Unsupervised Learning (UL), and Reinforcement Learning (RL). 

The SL algorithms can be classified into logistic regression \cite{xiao2017phy}, Linear Discriminant Analysis (LDA) \cite{wang2021safeguarding}, Decision Tree (DT) \cite{enad2020machine,pan2019threshold}, K-Nearest Neighbor (KNN) \cite{senigagliesi2022authentication}, Support Vector Machine (SVM) \cite{liu2017authenticating}, ensemble learning \cite{xie2021weighted,pan2019threshold,meng2023multiobservation}, extreme learning machine (ELM) \cite{wang2017physical}, Federated Learning (FL) \cite{wang2021collaborative}, and DL \cite{liao2019deep,chen2021physical}.
For instance, \cite{xiao2017phy} proposes a multi-landmark-based authentication scheme in which each landmark, equipped with multiple antennas, collects distributed Received Signal Strength Indications (RSSIs) to derive more distinctive characteristics. They utilize the Frank-Wolfe algorithm, grounded in logistic regression, to tackle the convex optimization problem associated with maximum likelihood-based coefficient estimation for the regression. \cite{enad2020machine} presents three ML-based PLA schemes, namely DT, SVM, and KNN. Simulations conducted on OFDM systems reveal that all ML-based approaches achieve greater detection accuracy compared to statistical-based methods. \cite{xie2021weighted} proposes an ensemble learning-based PLA approach tailored for edge computing environments. \cite{wang2021collaborative} develops a horizontal FL-based collaborative PLA scheme designed to alleviate the computational burden on IoT devices with limited processing and storage capabilities. The proposed distributed identification architecture assigns classification tasks to trustworthy collaborators, ultimately aggregating local parameters at a central device. \cite{chen2021physical} integrates TL with TP-Net to achieve lightweight and online identification.

The UL algorithms can be classified into clustering \cite{xia2021multiple,meng2023multidimensional}, one class classification-SVM (OCC-SVM) \cite{abdrabou2022adaptive}, manifold learning \cite{xia2022physical}, Gaussian Mixture Models (GMM) \cite{gulati2013gmm}, Gaussian process (GP) \cite{wang2021channel}, and AE \cite{meng2022physical}. 
For instance, \cite{xia2021multiple} investigates the relationship between multi-fingerprints and propose an unsupervised PLA scheme based on a non-parametric clustering algorithm. \cite{abdrabou2022adaptive} introduces an adaptive lightweight PLA approach that leverages antenna diversity techniques to enhance recognizable fingerprints. \cite{xia2022physical} applies manifold learning to tackle this issue by constructing a Markov chain of fingerprints in the time domain and evaluating the state transition probabilities of UAVs. \cite{gulati2013gmm} employs GMM to formulate the probabilistic model of transmitter radio channels, enabling online learning and parameter updates. \cite{wang2021channel} presents a PLA scheme based on Gaussian process channel prediction for IoT devices. Historical CSI fingerprints and the geographical data of devices are utilized to create a mapping that predicts the next legal CSI fingerprint for identification. \cite{meng2022physical} proposes a PLA scheme based on a hierarchical variational autoencoder to defend against spoofing attacks in IIoT, without requiring training fingerprints from attackers. The constructed loss function includes both an approximation and an exact calculation.

The RL algorithms can be classified into non-DL-aided RL \cite{gao2020physical} and Deep RL (DRL) \cite{lu2020reinforcement,wu2023game}.
For instance, \cite{gao2020physical} investigates a PLA scheme under the threat of intelligent spoofing attacks. The optimal transmit power allocation is derived to identify the optimal intelligent attack strategy for legitimate devices. \cite{wu2023game} examines three scenarios in static channels: multi-player games, zero-sum games with collisions, and zero-sum games without collisions. The closed-form expressions for Nash equilibrium are derived.

\section{DL-based PLA for Multi-Device Identification}
\label{Section III}
\begin{figure*}[h]
\centering
\includegraphics[width=0.95\textwidth]{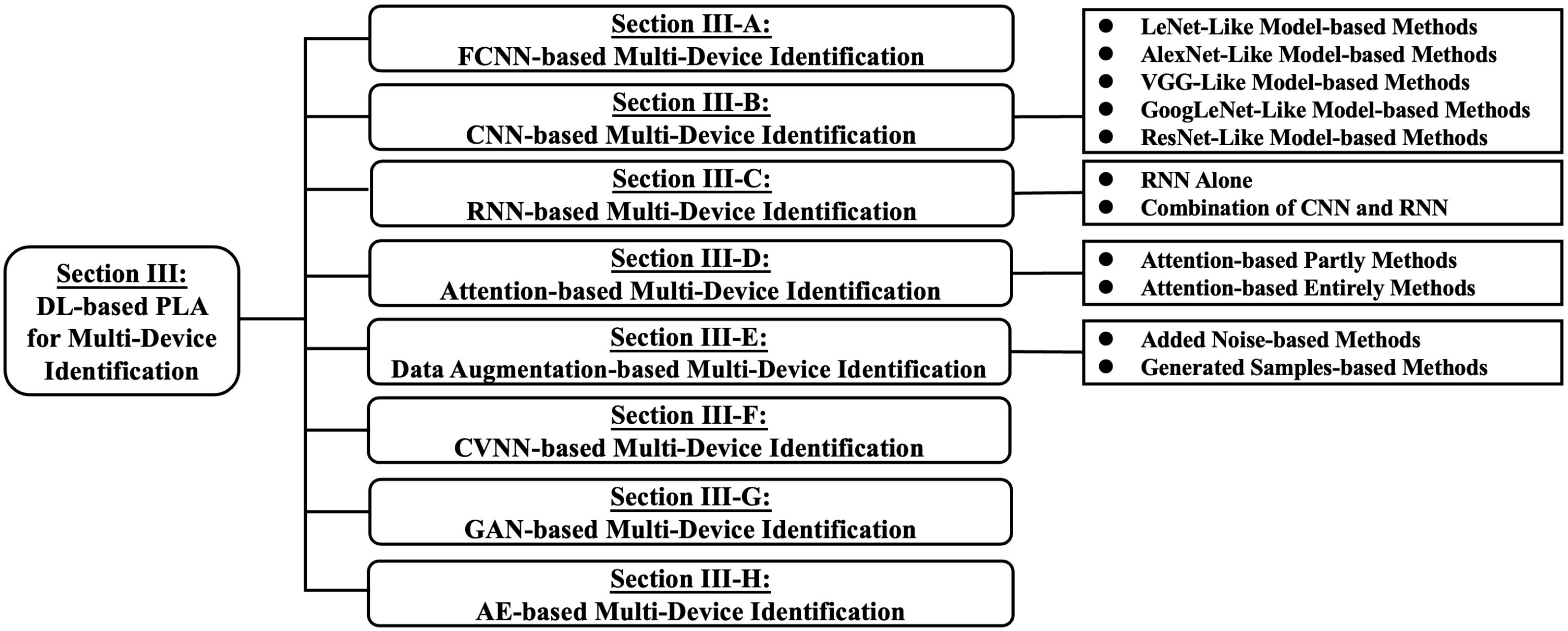}
\caption{Organization of Section III.}
\label{Organization of Section III}
\end{figure*}

In this section, we introduce the DL-based methods for multi-device identification. We divide the DL techniques into several sub-categories and compare them in detail. The organization of this section is illustrated in Fig. \ref{Organization of Section III}.

\subsection{FCNN-based Multi-Device Identification}
DL is a kind of ML techniques and is usually realized through neural networks, which are composed of one input layer, multiple hidden layers (also called latent layers), and one output layer. As illustrated in Fig. \ref{Illustration of typical DL techniques}, the FCNNs are directed layered neural networks without internal feedback connections. The latent layer $\bm{X}^{[i]}$ can be represented as
\begin{equation}
\bm{X}^{[i]}=\sigma(\bm{W}^{[i]}\bm{X}^{[i-1]}+\bm{\xi}^{[i]})
\end{equation}
where $\bm{W}^{[i]}$ and $\bm{\xi}^{[i]}$ respectively denote the weight matrix and bias matrix between the $(i-1)$th layer and the $i$th layer. $\sigma(\cdot)$ represents the activation function, including ReLU function, Tanh function, and Sigmoid function. In neural network, the activation function plays a crucial role. Its primary function is to introduce nonlinear characteristics, enabling neural networks to learn and simulate complex patterns. In a word, the activation function determines whether a neuron should be activated for a given input, thereby deciding what output should be produced.
\begin{itemize}
    \item \emph{Identification of USRP devices:} Roy et al. \cite{roy2019detection} design an FCNN-based scheme, where biases and regularization methods are utilized to avoid under-fitting and over-fitting issues. The Adam optimization algorithm is used to realize gradient descent and update network parameters. Roy et al. \cite{roy2019detection} further set up the laboratory testbed composed of several universal software radio peripheral (USRP) equipment and one RTL-Software Defined Radio (SDR) receiver to verify its effectiveness. The results reveal that the suggested FCNN-based scheme can identify 4 devices and 8 devices with the accuracy of 97.21\% and 96.6\%, respectively.
    \item \emph{Identification of ZigBee Devices:} Jafari et al. \cite{jafari2018iot} collect the I/Q fingerprints from 6 ZigBee devices (MICAz) and compare the authentication performance versus different slide window sizes and different learning rates.
\end{itemize}
\begin{les}
Although the FCNN-based schemes can obtain higher identification accuracy compared with the non-DL-based schemes, it is challenging for the fully-connected structure to extract the spatial information of fingerprints. Hence, the robustness needs to be enhanced, especially for large-scale fingerprint datasets \cite{agadakos2019deep}.
\end{les}

\subsection{CNN-based Multi-Device Identification}

\begin{figure*}[h]
\centering
\includegraphics[width=0.95\textwidth]{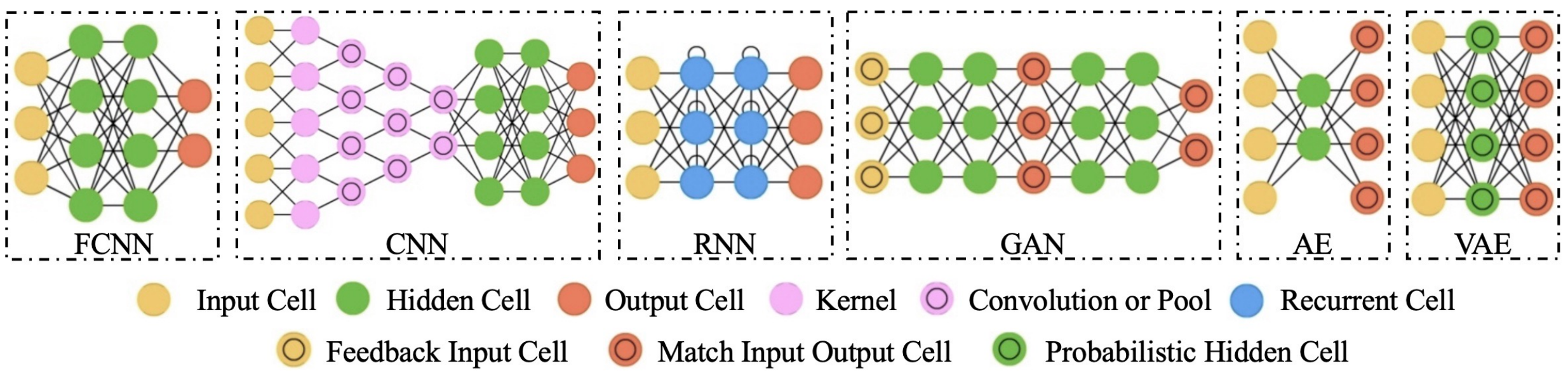}
\caption{Illustration of typical DL techniques, including FCNN, CNN, RNN, GAN, AE, and VAE.}
\label{Illustration of typical DL techniques}
\end{figure*}

To extract the spatial information of fingerprints, more and more researchers have resorted to CNNs. As illustrated in Fig. \ref{Illustration of typical DL techniques}, the latent layers in CNNs are usually composed of convolutional layers, pooling layers, and fully-connected layers. The output of the convolutional layer can be denoted as
\begin{equation}
\bm{X}^{[i]}=\sigma(\bm{W}^{[i]}\otimes\bm{X}^{[i-1]}+\bm{\xi}^{[i]})
\end{equation}
where $\bm{W}^{[i]}$  and $\bm{\xi}^{[i]}$  denote the convolution kernel and bias matrix between the $(i-1)$th layer and the $i$th layer, respectively. $\otimes$ is the convolution operation. The convolution operation defines a set of filters (also known as convolution kernels or feature detectors) that slide over the input image to extract specific local features such as edges, corners, and textures. By employing filters of varying sizes and shapes, CNNs can capture multi-scale and multi-level features. These extracted features are then combined and abstracted in subsequent layers of the network, resulting in a higher-level feature representation that is essential for complex image understanding tasks. The polling layer can compress data and reduce the number of parameters through parameter sharing. CNNs can realize the classification of pictures and are common in computer vision field. To obtain higher identification accuracy, fingerprints can be pre-processed as pictures to reserve spatial information, such as the input and output multi-antenna features of CSI fingerprints in MIMO systems and the in-phase components I and quadrature components Q of I/Q fingerprints. Here, we divide CNNs into five sub-categories as follows.
\subsubsection{LeNet-Like Model-based Methods}

\begin{figure*}[h]
\centering
\includegraphics[width=0.95\textwidth]{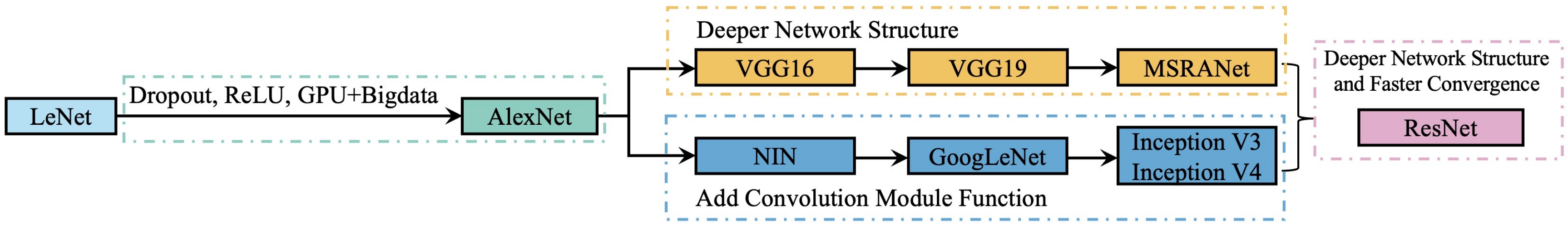}
\caption{Illustration of the evolutions of CNN models.}
\label{Illustration of the evolutions of CNN models}
\end{figure*}

As illustrated in Fig. \ref{Illustration of the evolutions of CNN models}, LeNet is one of the earliest CNNs and the origin of a large number of neural network architectures. LeNet is a lightweight CNN model and has been widely studied for PLA.
\begin{itemize}
    \item \emph{Radio Modulation Recognition:} Shea et al. \cite{o2016convolutional} first use the LeNet model to recognize radio modulation by raw I/Q fingerprints without expert transformation. The modulations for recognition include 8 digital modulations (BPSK, QPSK, 8PSK, 16QAM, 64QAM, BFSK, CPFSK, and PAM4) and 3 analog modulations (WB-FM, AM-SSB, and AM-DSB). The simulation results demonstrate that the CNN-based approach is feasible, especially at low SNR environments.
    \item \emph{Identification of Large-Scale Devices:} Sankhe et al. \cite{sankhe2019oracle} describe a CNN classifier named ORACLE to study the difference between I/Q samples of a large pool of bit-similar devices. The ORACLE architecture is consisted of 2 convolution layers and 2 fully-connected layers. Each complex I/Q sample, the input of the ORACLE model, is denoted as a 2-dimensional real value. Sankhe et al. \cite{sankhe2019oracle} utilize the 16-node USRP X310 SDR testbed to generate I/Q fingerprints and use 140 COTS WiFi devices to obtain external datasets. The authentication results reveal that ORACLE can realize $\ge 99\%$ identification accuracy on static radio environments.
    \item \emph{Identification of Mobile Phones:} Aneja et al. \cite{aneja2018iot} set up the experiment system with iPhone 7 Plus and iPad 4 and use inter arrival time of the identifying signatures. The results indicate that the CNN classifier can realize 86.7\% accuracy.
    \item \emph{Performance versus Different Window Sizes:} Jafari et al. \cite{jafari2018iot} test the authentication performance of the CNN model on ZigBee devices. The results reveal that the simulation experiments verify that the CNN model can realize the accuracy of 94.7\%, 94.2\%, and 94\% with the window size of 32, 64, and 128, respectively.
\item \emph{CNNs Combined with Clustering:} Wong et al. \cite{wong2018clustering} investigate the CNN model as the characteristic extractor, paired with density-based spatial clustering of applications with noise \cite{ester1996density} clustering algorithm. The introduction of the clustering algorithm is beneficial for the visualization of fingerprints and better understanding of the further authentication process. The simulation results indicate that when there are 25 transmitters, the proposed model can achieve 78.7\% and 80.4\% accuracy with 0.25 MHz and 1.67 MHz bandwidth, respectively.
\item \emph{Performance versus Different SNRs:} Merchant et al. \cite{merchant2018deep} verify the performance of the CNN architecture on a set of IEEE 802.15.4 devices. The fingerprints are collected from 7 DigiXBP24CZ7SITB003 ZigBee Pro devices with the 204 GHz band, and the CNN technique is demonstrated via two experiments: a noisy channel and a lab environment with simulated additive white Gaussian noises. The results show that when SNR is 10 dB and 40 dB, the accuracy can reach 73.73\% and 91.38\%, respectively.
\item \textit{Lightweight Realization:} Qi et al. \cite{qi2024lightweight} focus on intercepting and representing signals to address concerns of complexity and efficiency. A comprehensive complexity analysis of PLA models is conducted, including parameters, floating-point operations, and memory usage.

\end{itemize}
\subsubsection{AlexNet-Like Model-based Methods}
As illustrated in Fig. \ref{Illustration of the evolutions of CNN models}, compared with LeNet, AlexNet exploits ReLU function as the activation function to obtain better performance in deep network architectures and use Dropout to address the over-fitting issue. In addition, AlexNet can leverage GPU to realize more powerful computational ability. For these reasons, AlexNet-enabled PLA schemes are more efficient.
\begin{itemize}
    \item \emph{Performance under Noisy Multipath Environments:} Riyaz et al. \cite{riyaz2018deep} propose a CNN-based identification architecture that is motivated in part by AlexNet. Riyaz et al. \cite{riyaz2018deep} further used 720,000 fingerprints, 80,000 fingerprints, and 200,000 fingerprints for training, validation, and testing, respectively. The simulation results reveal that the CNN model can achieve 98\% identification accuracy for 5 devices on the noisy multipath wireless channels.
\item \emph{Performance on ADS-B Datasets:} Shawabka et al. \cite{al2020exposing} collect I/Q datasets with 7 TB from 20 transmitters with WiFi and automatic-dependent surveillance-broadcast (ADS-B) transmissions and discuss the impact of wireless channels on the identification accuracy. The results indicate that balancing I/Q samples can significantly enhance the identification accuracy when there are a large number of transmitters.
\item \emph{Key-Aided Identification:} Sankhe et al. \cite{sankhe2019no} provide a CNN architecture with 8 convolution layers and 2 fully-connected layers for large-scale fingerprint datasets. Sankhe et al. \cite{sankhe2019no} further present a scalable identification approach named impairment hopping spread spectrum, which can generate random binary sequence keys to defend against spoofing attacks.
\item \emph{Study on Interpretability:} Liu et al. \cite{liu2020zero} pay attention to enabling DL techniques to be practically dependable in device identification and propose to employ the zero-bias layer in the CNN model to realize interpretable and dependable identification. The simulations on real-world ADS-B transmitters demonstrate its effectiveness.
\item \emph{CNNs Combined with Metric Learning:} To achieve lightweight identification without relying on a large number of samples, Gaskin et al. \cite{gaskin2022tweak} present Tweak technique that can create a portable model to identify LoRa devices. The proposed Tweak method leverages metric learning to effectively process the fingerprints from non-original training domains.
\item \emph{Impact of Receivers on Performance:} Shen et al. \cite{shen2022towards161} study the impact of receivers on identification performance and propose an adversarial training approach. They further provide two training strategies, including homogeneous and heterogeneous training. The simulation results verify that it can obtain over 75\% accuracy for 20 SDRs.
\item \emph{CNNs Combined with Ensemble Learning:} Huang et al. \cite{huang2022radio} combine CNN models and ensemble learning to propose two identification architectures, including the improved CNN by Bagging and the improved CNN by Boosting. The simulations verify that the proposed method can achieve the accuracy of 89.75\% in the 10 dB noise scenarios.
\item \emph{Channel Robust Identification:} Xing et al. \cite{xing2022design} propose a channel robust identification approach named the difference of the logarithm of the spectrum. The proposed method is independent on a single fingerprint data and does not require additional manipulation of the transmitters. The simulations in the IEEE 802.11 OFDM system show that the proposed scheme can reach the accuracy of 99.02\% and 97.05\% in the single- and cross- environment evaluations, respectively.

\end{itemize}

\subsubsection{VGG-Like Model-based Methods}
As illustrated in Fig. \ref{Illustration of the evolutions of CNN models}, compared with AlexNet, VGG has a deeper network structure. The convolution operations are achieved through multiple stacked convolution kernels with a small size, which indicates that it requires fewer parameters and has better feature extraction ability. Hence, VGG-based PLA methods are more suitable for lightweight deployment.
\begin{itemize}
    \item \emph{Study on In-Band and Out-of-Band Spectrum Emissions:} Elmaghbub et al. \cite{elmaghbub2020leveraging} consider both the in-band and out-of-band spectrum emissions to extract the fingerprints, which can provide discriminate identifying signatures even when the distortion of hardware of different devices is significantly reduced. The experiments show that the accuracy for 5 devices can reach 96.2\% whereas that of the in-band-based method is 48.6\%.
\item \emph{Verification of VGG-16:} Zong et al. \cite{zong2020rf} propose an authentication architecture based on VGG-16 and use the fingerprints of 5 transmitters with SNR of 20 dB to verify its effectiveness.
\item \emph{CNNs Combined with TL} Chen et al. \cite{chen2020physical} develop a CNN architecture named TP-Net to realize effective decision-making process at the edge devices. The simulations on Rayleigh models demonstrate its superiority in accuracy than CNN-2, GC-Net, and VGG-7. Chen et al. \cite{chen2021physical} further combine TL and TP-Net to realize lightweight and online identification for 40 devices.

\end{itemize}
\subsubsection{GoogLeNet-Like Model-based Methods}
In GoogLeNet, the basic convolution block is called inception block, whose core concept is to use the combination of different sizes of convolution kernels to fuse various information of different scales. Such design can deepen the network without increasing the number of parameters. 
\begin{itemize}
    \item \emph{Comparison with AlexNet:} Zha et al. \cite{zha2020real} verify the accuracy performance of AlexNet and GoogLeNet on 1090MHz baseband signals collected from 5 aircraft. The results under different SNRs demonstrate the effectiveness of the contour stellar images.
\item \emph{Identification under FHSS Networks:} Kang et al. \cite{kang2021radio} propose the RFEI algorithm based on Inception-A for FHSS networks. The results show that the proposed approach can realize classification accuracy of 97.0\% for 7 unseen real FHSS transmitters.
\item \emph{Comparison with CNN-3, VGG-16, and ResNet-50:} Mcmillen et al. \cite{mcmillen2023deep} propose an RF fingerprint system named PUWS based on chaotic antenna arrays. A basic Convolutional Neural Network (CNN) with two convolutional layers, each containing 64 neurons, followed by a single dense layer, is employed as the baseline for performance comparison among the models discussed. These models include VGG-16, a 16-layer neural network; ResNet-50, which features 50 layers and introduces residual connections between layers; Inceptionv3, a deep CNN that utilizes a 'network-in-network' strategy to learn features more profoundly; and Xception, an Inception-based model that enhances accuracy through the use of residual connections and separable convolutional layers. Initially designed for image classification with typical inputs of size 224 × 224 × 3, these models required adjustments to their top layers to handle received I/Q samples of size 1000 × 8 × 1, where the 8 columns represent I and Q signal samples from 4 antenna elements. In comparison to CNN-3 (with an accuracy of 93.3\%), VGG-16 (93.5\%), and ResNet-50 (99.2\%), the proposed Inception and Xception models achieve an accuracy of 99.9\% for 300 devices.

\item \emph{Verification of MSCNN:} Based on the convolution layers of multiple branches with different convolution kernel sizes, Zhang et al. \cite{zhang2022specific} develop an MSCNN to enhance the identification performance. The proposed MSCNN model can exploit the inherent features in multiple receptive fields. The simulations under perfect environments verify that the proposed MSCNN method can enhance the absolute accuracy and relative accuracy by 15\% and 22\%, respectively.
\end{itemize}

\subsubsection{ResNet-Like Model-based Methods}
Through residual learning, ResNet can address the gradient explosion/ disappearance issues caused by the deep structure. ResNet is suitable for the feature extraction of large-scale datasets and can obtain more robust identification models.
\begin{itemize}
\item \emph{Identification of Satellite Transmitters:} Oligeri et al. \cite{oligeri2022past} introduce PAST-AI to realize the PLA of satellite transmitters via DL techniques. They provide two device identification scenarios: intra-constellation satellite identification and satellite identification in the wild, and further compare the accuracy and training overhead of ResNet-18 and baseline models.
\item \emph{Scalability Analysis for Large-Scale Devices:} Jian et al. \cite{jian2020deep} analyze the scalability issue on identifying very large device populations, including 10,000 devices, and further present comprehensive performance comparison of different CNN models, channel environments, SNR, number of devices, and dataset size.
\item \emph{Identification without Retraining:} Gritsenko et al. \cite{gritsenko2019finding} develop an approach to identify new devices without retraining a classifier, and provide comprehensive analysis of the designed architecture from the perspective of model parameters and I/Q datasets. Gritsenko et al. \cite{gritsenko2019finding} test the technique on 6 real-world datasets with different sizes and transmission protocols. The proposed approach can detect unseen devices with the accuracy of 76\%, while reducing the identification accuracy of 500 previously seen devices by no more than 10\%.
\item \emph{Identification of Mobile Phones:} Zhang et al. \cite{zhang2021deep171} propose a ResNet-based approach named RFFResNet to authenticate real mobile phones, and provide the influence of channel environments, noises, training dataset scale, and model parameters. The performance of RFFResNet is tested on the LTE simulation datasets with 220 GB and real mobile phone’s datasets with 25 GB. The results verify that RFFResNet can realize 95\%-99\% accuracy, which is higher than that of ResNet18-1D, ResNet34-1D, and VGG16-1D.
\item \emph{Identification under Low SNR Environments:} Tang et al. \cite{tang2021specific} design a DRSN to identify devices in low SNR environments. The soft threshold is utilized to preserve more useful characteristics, and the identity shortcut is employed to improve the training speed. The simulations verify the superiority of the proposed DRSN architecture in accuracy than FCNN \cite{jafari2018iot}, CNN 1D \cite{merchant2018deep}, CNN 2D \cite{yu2019robust}, and ResNet \cite{pan2019specific}.
\item \emph{CNNs Combined with Structured Pruning:} To enhance the identify accuracy, Jian et al. \cite{jian2021radio} propose the ResNet50-1D architecture, containing 49 convolutional layers and one fully-connected layer. They utilize structured pruning \cite{zhang2018systematic} to generate the compressed versions efficiently and use an ADMM approach to prune the model \cite{ren2019admm}. Numerous experiments on multi-edge-devices verify the high-efficiency of the proposed method.

\end{itemize}

Here, we summarize the contributions of all CNN-based multi-device identification schemes in Tab. \ref{CNN-based Multi-Device Identification Schemes}.

\begin{longtable}{|A{2cm}|A{1cm}|A{1cm}|P{7.6cm}|}
\caption{CNN-Based Multi-Device Identification Schemes} \label{CNN-based Multi-Device Identification Schemes} \\
\hline
\textbf{CNN} & \textbf{Ref.} & \textbf{Year} & \textbf{\begin{tabular}[c]{@{}c@{}}Major Contribution\end{tabular}} \\ \hline
\endfirsthead 
\hline
\textbf{CNN} & \textbf{Ref.} & \textbf{Year} & \textbf{\begin{tabular}[c]{@{}c@{}}Major Contribution\end{tabular}} \\
\hline
\endhead

\hline
\endfoot

\hline
\endlastfoot

\hline
\multirow{7}{*}[-15ex]{\begin{tabular}[c]{@{}A{2cm}@{}} LeNet-Like Models \end{tabular}}    & \cite{sankhe2019oracle}  & 2019 & \begin{tabular}[c]{@{}P{7.6cm}@{}}Propose ORACLE, a LeNet-based classifier, to realize the identification for a large pool of bit-similar devices.\end{tabular} \\
\cline{2-4} & \cite{aneja2018iot} & 2019 & Used IAT as identifying signatures and verify the performance of the CNN classifier for iPhone 7 Plus and iPad4. \\
                      \cline{2-4}
                      & \cite{o2016convolutional} & 2016 & First utilize LeNet to recognize radio modulation by raw I/Q fingerprints without expert transformation. \\
                      \cline{2-4}
                      & \cite{jafari2018iot} & 2019 & Test the identification performance of CNNs for ZigBee devices versus different window sizes. \\
                      \cline{2-4}
                      & \cite{wong2018clustering} & 2019 & Combine CNNs and spatial clustering algorithm to realize the visualization of fingerprints. \\
                      \cline{2-4}
                      & \cite{merchant2018deep} & 2018 & Verify the performance of CNNs for a set of IEEE 802.15.1 devices in noisy channels and a lab environment, respectively.

\\
\cline{2-4}
& \cite{qi2024lightweight} & 2019 &Analyze the complexity of PLA models, including parameters, floating-point operations, and memory usage.\\
\hline
\multirow{8}{*}[-15ex]{\begin{tabular}[c]{@{}A{2cm}@{}} AlexNet-Like Models \end{tabular}}  & \cite{riyaz2018deep}  & 2018 & Provide an AlexNet-based classification architecture and verify the performance using 1,000,000 fingerprints.                                                                                                                                \\
\cline{2-4}
                      & \cite{al2020exposing} & 2020 & Collect I/Q datasets with 7 TB from 20 transmitters with WiFi and ADS-B transmissions and discuss the impact of channels on accuracy.                                                                                                        \\
                      \cline{2-4}
                      & \cite{sankhe2019no} & 2020 & Present a CNN architecture for large-scale fingerprint datasets and further present a scalable method for attack detection.                                                                                                                  \\
                      \cline{2-4}
                      & \cite{liu2020zero} & 2021 & Employ the zero-bias layer in CNNs to achieve interpretable and dependable identification.                                                                                                                                                    \\
                      \cline{2-4}
                      & \cite{gaskin2022tweak} & 2022 & Leverage metric learning and CNNs to effectively process the fingerprints from non-original training domains.                                                                                                                              \\
                      \cline{2-4}
                      & \cite{shen2022towards161} & 2022 & Study the impact of receivers on performance and propose an adversarial training method, including homogeneous and heterogeneous training.                                                                                                  \\
                      \cline{2-4}
                      & \cite{huang2022radio} & 2022 & Combine CNNs and ensemble learning to propose two methods, including the improved CNNs by Bagging and Boosting, respectively.  \\
                      \cline{2-4}
                      & \cite{xing2022design} & 2022 & Show a channel robust identification approach, which is independent on a single fingerprint data and does not require additional manipulation of transmitters.                                                                               \\
\hline
\multirow{3}{*}[-8ex]{\begin{tabular}[c]{@{}A{2cm}@{}} VGG-Like Models  \end{tabular}}      & \cite{elmaghbub2020leveraging} & 2020 & Consider both the in-band and out-of-band spectrum emissions and provide a method that can provide discriminate identify signatures even when the distortion of hardware of different devices is significantly reduced.  \\ 
\cline{2-4}
                      & \cite{zong2020rf} & 2020 &  Develop an architecture based on VGG-16 and confirm its performance on 5 transmitters with SNR of 20 dB. \\
                      \cline{2-4}
                      & \cite{chen2021physical} & 2022 &Propose TP-Net based on CNNs to realize effective decision-making process at the edge devices, and                                        \\
                      & \cite{chen2020physical} & 2020 & further combine TL and TP-Net to realize lightweight and online identification for 40 devices.                                                                                                                                                                                                                                              \\
                      \hline
\multirow{4}{*}[-5ex]{\begin{tabular}[c]{@{}A{2cm}@{}} GoogLeNet-Like Models  \end{tabular}}   & \cite{zha2020real} & 2020 &  Compare the performance of AlexNet and GoogLeNet on 1090 MHz baseband signals collected from 5 aircraft.\\
                      \cline{2-4}

                      & \cite{kang2021radio} & 2021 & Introduce the RFEI method based on Inception-A for FHSS networks and verify the performance on 7 FHSS transmitters.                                                                                                                      \\
                      \cline{2-4}
                      
                      & \cite{mcmillen2023deep} & 2023 & Present an authentication framework based on CAAS and compare the performance of CNN-3, VGG-16, ResNet-50, Inception, and Xception. \\
                      \cline{2-4}
                      & \cite{zhang2022specific} & 2022 & Develop a multi-scale-CNN architecture to exploit the inherent features in multiple receptive fields and enhance accuracy.\\
                      \hline
                      
\multirow{6}{*}[-8ex]{\begin{tabular}[c]{@{}A{2cm}@{}} ResNet-Like Models \end{tabular}}      & \cite{oligeri2022past}   & 2023 & Propose PAST-AE to achieve the PLA of satellite transmitters under two typical scenarios: intra-constellation satellite identification and satellite identification in the wild.  \\
                      \cline{2-4}
    & \cite{jian2020deep}  & 2020 & Analyze the scalability issue of applying identification methods to very large device populations, and further present comprehensive performance comparison of different CNN models, channel environments, number of devices and dataset size. \\
                      \cline{2-4}
                      
                      & \cite{gritsenko2019finding} & 2019 &Present a method without retraining classifiers, and provide comprehensive analysis from the perspective of model parameters and I/Q datasets. \\
                      \cline{2-4}& \cite{zhang2021deep171} & 2021 & Propose RFFResNet to identify real mobile phones, and compare their performance with RstNet18-1D, ResNet34-1D, and VGG16-1D.\\
                      \cline{2-4}
                      
                      & \cite{tang2021specific} & 2021 &Design DRSN to identify devices in low SNR environments, and compare its performance with FCNN, CNN 1D, CNN 2D, and ResNet. \\
                      \cline{2-4}
                      
                      & \cite{jian2021radio} & 2022 & Propose ResNet50-1D architecture that utilizes structured pruning to generate the compressed versions  efficiently and uses ADMM trick to prune models. 
                      \\
\hline 
\end{longtable}

\begin{les}
CNNs are the most widely used models to extract the spatial information of RF fingerprints and channel fingerprints, especially I/Q fingerprints and CSI fingerprints, respectively. The real part and imaginary part of complex signals correspond to different channels of CNNs. Hence, fingerprints are pre-processed as pictures to further extract the frequency domain information. For the LeNet-like and AlexNet-like models, the combinations with clustering algorithms \cite{wong2018clustering}, metric learning \cite{gaskin2022tweak}, and ensemble learning \cite{huang2022radio} as well as the performance for mobile phones \cite{aneja2018iot}, noisy multipath environments \cite{riyaz2018deep}, WiFi scenarios \cite{al2020exposing}, ADS-B scenarios \cite{al2020exposing}, and ZigBee devices \cite{jafari2018iot} have been studied. However, due to the limited network architecture, it is challenging for them to identify large-scale devices. For the VGG-like models, the network structure is deeper, and TL is combined to realize lightweight and online identification for 40 devices \cite{chen2021physical}. For the GoogLeNet-like models, the introduction of the inception block can deepen the network without increasing the number of parameters. Such design enables the identification to extend to large-scale scenarios with 300 devices \cite{mcmillen2023deep}. For the ResNet-like models, residual learning involved is suitable for the feature extraction of large-scale datasets, such as the very large device populations including 10,000 devices \cite{jian2020deep} and the LTE simulation datasets with 220 GB \cite{gritsenko2019finding}.
\end{les}

\subsection{RNN-based Multi-Device Identification}

\begin{figure*}[h]
\centering
\includegraphics[width=0.95\textwidth]{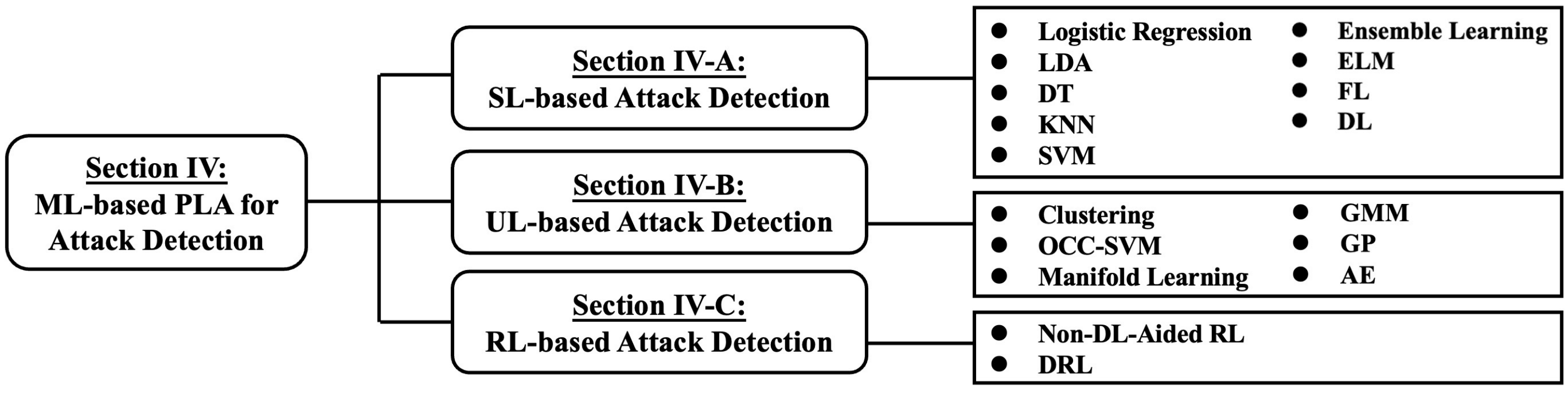}
\caption{Organization of Section IV.}
\label{Organization of Section IV}
\end{figure*}

As illustrated in Fig. \ref{Organization of Section IV}, RNN can extract the time series characteristics of fingerprints through recursion operations. Typical RNN models include Long Short-Term Memory (LSTM) and Gated Recurrent Unit (GRU).
\subsubsection{RNN Alone}
\begin{itemize}
\item \emph{Comparison between FCNN, CNN, and RNN:} Jafari et al. \cite{jafari2018iot} compare the identification performance between FCNN, CNN, and RNN. The simulations on I/Q samples collected from ZigBee devices over several SNR levels show the inferiority of RNN.
\item \emph{Performance under Low SNR Environments:} Wu et al. \cite{wu2018deep} propose a LSTM-based approach to capture the long-term and short-term characteristics of I/Q fingerprints. The experimental studies under low SNR environments (-12 dB) demonstrate its effectiveness.
\item \emph{Novel Metrics:} Shawabka et al. \cite{al2021deeplora} assess the performance of CNN (1D CNN and 2D CNN) and LSTM utilizing the following metrics: “Per-slice Training” accuracy, “Per-slice Testing” accuracy, “Train-and-Test-Same-Day” accuracy, and “Train-and-Test-Other-Day” accuracy. Massive experiments verify the superiority of CNN than RNN in identification performance.
\end{itemize}

\subsubsection{Combination of CNN and RNN}
Considering that the combination of CNN and RNN can extract the spatial and temporal characteristics of fingerprints, researchers have conducted a series of studies.
\begin{itemize}
\item \emph{Identification of Large-Scale WiFi Devices:} Soltani et al. \cite{soltani2020more} provide the RFF-ConvRNN model, consisting of the convolution module, RNN module, and fully-connected module for extracting spatial properties, extracting time-dependent features, and classification, respectively. The simulations on 50 WiFi devices indicate that when data augmentation is not used, the proposed RFF-ConvRNN model can obtain better identification performance than baseline models.
\item \emph{Performance under Low SNR Environments:} Liu et al. \cite{liu2020specific} suggest the DBi-LSTM and the one-dimensional residual convolutional network with dilated convolution and squeeze-and-excitation block to against the interference of unreliable fingerprints. The combination of DBi-LSTM and CoBONet can exploit the fine-grained attributes of signals and directly extract the temporal characteristics from baseband I/Q fingerprints. The simulations on the open-source datasets \cite{sankhe2019oracle} collected from 16 USRP devices show that the proposed architecture can obtain high accuracy even in low SNR environments.
\item \emph{Comparison with LSTM and GRU:} Roy et al. \cite{roy2019rf} combine LSTM and CNN to exploit the temporal and spatial properties of I/Q fingerprints. The I/Q samples are generated from 8 different USRP B210s under an indoor environment with the SNR of 30 dB. The simulation results show that the proposed ConvLSTM can realize the accuracy of 97.2\%, while LSTM and GRU can reach the accuracy of 92\% and 95.3\%, respectively.

\end{itemize}
\begin{les}
Due to the insufficient extraction of spatial information, the utilization of RNNs alone can not obtain higher identification performance than CNNs. One feasible approach is to combine CNNs and RNNs to extract both temporal and spatial properties. Such design has been verified in large-scale WiFi scenarios \cite{soltani2020more} and low SNR environments \cite{liu2020specific}.
\end{les}

\subsection{Attention-based Multi-Device Identification}
The Attention mechanism refers to selectively paying attention to a part of all information while ignoring other visible information. The Attention mechanism can help efficiently extract useful features.
\subsubsection{Attention-based Partly Methods}
\begin{itemize}
\item \emph{Attention to Frequency Domain:} Considering that the ability of feature extraction in time domain is limited, Peng et al. \cite{peng2021specific} propose a squeeze-and-excitation neural network (SeNet) approach in frequency domain. The suggested architecture can realize channel attention mechanism using the squeeze-and-excitation module so that the useful characteristics of fingerprints are emphasized and the useless information is weakened. The simulations on 20 WiFi devices verify that the proposed SeNet can obtain better identification performance and stronger robustness than baseline models even in low SNR environments.
\item \emph{CNNs Combined with Attention:} To leverage rough priori information to improve authentication performance, Weng et al. \cite{weng2020message} pay attention to message structure and provide the MSCAN. The proposed model can separate the portions with different pulse waveform distributions and achieve spatial attention mechanism for low-dimensional fingerprints. Such design is beneficial to the exploration of internal correlation and more efficient feature fusion. The simulation results on fingerprints collected from real-world ADS-B transmissions show that the suggested MSACN can reach the accuracy of 98.20\% for 44 transmitters.
\item \emph{Comparison with AlexNet, VGG, and AttMsCN:} Zhang et al. \cite{zhang2023data} propose a data-and-knowledge dual-driven architecture, including the multi-scale module, attention module, and classifier. The proposed model can exploit more useful semantic information from protocol knowledge and explore higher-level characteristics using the attention mechanism. The simulations on 5 WiFi devices show that the accuracy of the proposed model is 55\%, 44\%, and 35\% higher than VGG, CNN \cite{riyaz2018deep}, and AttMsCN with the SNR of -10 dB.
\end{itemize}
\subsubsection{Attention-based Entirely Methods (Transformer)}
\begin{itemize}
\item \emph{Performance under Low SNR Environments:} Shen et al. \cite{shen2021radio114} propose a transformer model to identify LoRa devices with the signals of variable lengths and propose a multi-packet inference approach to greatly enhance the accuracy performance in low SNR environments. The simulations on 10 LoRa devices show that the proposed method can improve the identification accuracy from 60\% to 90\% with the SNR of 10 dB.
\item \emph{Identification of Unknown Devices:} To handle devices that appear outside the closed set, Xu et al. \cite{xu2021transformer} provide a novel identification approach combining an enhanced Transformer and a modified ICS algorithm. The proposed architecture can recognize unseen classes of devices while maintaining good identification performance of known devices. The simulations on 30 distinct devices show that the proposed model is superior in identification accuracy and robustness than baseline models, including Hybrid OvA \cite{hanna2020open}, Modified ICS \cite{xu2020open}, MLOSR \cite{oza2019deep}, and CROSR \cite{yoshihashi2019classification}.
\end{itemize}
\begin{les}
The Attention mechanism is usually combined with CNNs to emphasize the useful characteristics of fingerprints and weaken the useless information. In addition, some researchers study the identification methods based on Attention mechanism entirely, that is, Transformer. Transformer uses encoders and decoders to extract features. Compared with CNNs and RNNs, Transformer has a global vision and better parallel computing ability as well as can handle variable-length fingerprints. Hence, it has better flexibility and scalability. However, the training of Transformer-based identification methods requires more training fingerprints, limiting its application.
\end{les}
\subsection{Data Augmentation-based Multi-Device Identification}
The training of DL-based identification methods usually requires many fingerprint samples, and insufficient fingerprints will cause the over-fitting issue, thus degrading the identification accuracy. To address this issue, one promising approach is data augmentation, which is divided into two sub-categories: added noise-based and generated samples-based methods. The former one leverages noises to improve the generalization of identification models, while the latter one improves the robustness of fingerprints by generating synthetic fingerprint samples.
\subsubsection{Added Noise-based Methods}
\begin{itemize}
\item \emph{Online Augmentation:} Shen et al. \cite{shen2021radio114} propose a Gaussian noise-based enhancement method and test the identification accuracy of the Transformer-based classifier with online-augmentation-based training, offline-augmentation-based training, and non-augmentation-based training. The results reveal that the online-augmentation-based classifier can obtain higher accuracy.
\item \emph{Noises Added to Latent Layers:} Different from adding Gaussian noises in the input layer, Meng et al. \cite{meng2022multiuser} proposed a LPNN approach to guarantee the access security of IIoT. Due to the latent layer has stronger linear characteristics than the input layer for CSI fingerprints, the proposed LPNN method has better interpretability. Meng et al. \cite{meng2022multiuser} further define Fingerprint Library to provide the post-hoc explanations of the authentication system. The simulations under the static and dynamic IIoT environments demonstrate the superiority in identification accuracy of the proposed LPNN model than vanilla models.
\item \emph{Channel Fading Models Combined with Noises:} Shen et al. \cite{shen2022towards} leverage channel fading models and add the artificial white Gaussian noise to the signal to emulate different SNR levels to realize data augmentation. The fading models consider both the multipath and Doppler effect. The proposed approach can improve the identification accuracy in high-speed scenarios from 68.6\% to more than 80\%.
\item \emph{Performance under Dynamic Environments:} Zhang et al. \cite{zhang2023data187} propose a device identification system, including the registration and detection stages. In the registration, the artificial noise with random power is added into the training fingerprints to improve the channel robustness. The simulations under the real-world environments, the laboratory static environments, and the laboratory dynamic environments have demonstrated its feasibility.
\end{itemize}
\subsubsection{Generated Samples-based Methods}
\begin{itemize}
\item \emph{Data Augmentation Combined with Slicing Technique:} Yu et al. \cite{yu2019robust} design a multi-sampling-CNN to identify ZigBee devices and utilize the slicing techniques to obtain more fingerprint samples. The proposed slicing techniques can avoid the over-fitting issues effectively and solve the timing error during low SNR environments.
\item \emph{Comparison between Different Data Augmentation Methods:} Huang et al. \cite{huang2019data} evaluate different data augmentation approaches for the DL-aided modulation classification, including rotation, flip, and Gaussian noise. The simulations on the open-source datasets, RadioML2016.10a, verify that all three approaches mentioned above can enhance the classification accuracy and accelerate the training speed, but the Gaussian noise-based method is worse than the other methods. Zhang et al. \cite{zhang2022data} compare four data augmentation methods for the DL-based device identification, including flip, rotation, shift, and noise. The simulations on few-shot ADS-B verify their effectiveness. The proposed approaches can be extended to different classes of few-shot fingerprints.
\item \emph{Data Augmentation Combined with Random Integration:} Xie et al. \cite{xie2019data} study the correlation between RF fingerprints and deoxyribonucleic acid sequency and design a random integration-based data augmentation approach. The proposed pseudo-random integration can obtain higher accuracy than traditional approaches and the random integration method.
\item \emph{Data Augmentation Combined with FL:} To guarantee the performance of DL-based identification techniques in time-varying channels, Piva et al. \cite{piva2021tags} leverage federated data augmentation method to propose a novel training structure. The simulation results on 200 SDRs verify that the proposed approach can enhance the accuracy of the FL-based scheme by up to 19\%.
\item \textit{Adversarial Data Augmentation:} Liu et al. \cite{liu2024overcoming} employ self-supervised learning during the pre-training phase to mitigate the reliance on labeled fingerprint samples. In the fine-tuning phase, knowledge transfer is utilized to alleviate the dependence on samples from the target dataset. Additionally, by introducing a “relaxation” mechanism in the feature space, an adversarial augmentation method that facilitates flexible and dynamic data transformations is further proposed, thereby enabling effective self-supervised learning.

\item \emph{Performance under Industrial Scenarios:} Liao et al. \cite{liao2019multiuser} propose three data augmentation methods to generate additional CSI samples for the training of the FCNN-based authentication model, including the averaging data augmentation, the exponential averaging data augmentation, and the stochastic weight averaging data augmentation. The simulations on real industrial datasets demonstrate that the proposed methods can accelerate the training speed and enhance the authentication accuracy.
\end{itemize}
\begin{les}
The added noise-based data augmentation methods include adding noises in the input layers \cite{shen2021radio114,zhang2023data187} and latent layers \cite{meng2022multiuser} as well as combining noises with channel fading models \cite{shen2022towards}. The generated samples-based methods include typical ration, flip, and shift \cite{huang2019data,zhang2022data} as well as integration methods \cite{xie2019data} and generative models \cite{roy2019detection,karunaratne2021open}. Moreover, the combination of different data augmentation methods may obtain better performance.
\end{les}

Here, we summarize the contributions of all multi-device identification schemes based on RNN, Attention, and data augmentation in Tab. \ref{Multi-Device Identification Schemes}.
\newpage
\begin{longtable}{|A{1.5cm}|A{2cm}|A{1cm}|A{1cm}|P{6.1cm}|}
\caption{Multi-Device Identification Schemes based on RNN, Attention, and Data Augmentation} \label{Multi-Device Identification Schemes} \\
\hline
\multicolumn{2}{|c|}{\textbf{Models}}& \textbf{Ref.} & \textbf{Year} & \textbf{\begin{tabular}[c]{@{}P{6.1cm}@{}}Major Contribution\end{tabular}} \\ \hline
\endfirsthead 
\hline
\multicolumn{2}{|c|}{\textbf{Models}} & \textbf{Ref.} & \textbf{Year} & \textbf{\begin{tabular}[c]{@{}P{6.1cm}@{}}Major Contribution\end{tabular}} \\
\hline
\endhead

\hline
\endfoot

\hline
\endlastfoot

\hline

\multirow{6}{*}[-10ex]{\begin{tabular}[c]{@{}A{1.5cm}@{}} RNN \end{tabular}} & \multirow{3}{*}[-6ex]{\begin{tabular}[c]{@{}A{2cm}@{}}RNN Alone\end{tabular}}  & \cite{jafari2018iot} & 2019 & Compare the performance between FCNN, CNN, and RNN on ZigBee devices  \\ \cline{3-5} 
 &  & \cite{wu2018deep} & 2018 & Propose a LSTM-based method to capture the long-term and short-term features of I/Q fingerprints. \\ \cline{3-5} & & \cite{al2021deeplora} & 2021 & Assess the identification performance of CNN and LSTM through massive experiments. \\ \cline{2-5} & \multirow{3}{*}[-6ex]{\begin{tabular}[c]{@{}A{2cm}@{}} Combination of CNN and RNN \end{tabular}} & \cite{soltani2020more} & 2020 & Present RFF-ConvRNN to extract spatial and time-dependent features.\\ \cline{3-5}  & \multicolumn{1}{l|}{}  & \cite{liu2020specific} & 2020 & \begin{tabular}[c]{@{}P{6.1cm}@{}}Provide DBi-LSTM and Conv-OrdsNet to against the interference of unreliable fingerprints, and verify the performance on the open-source datasets \cite{sankhe2019oracle}.\end{tabular} \\ \cline{3-5}  & \multicolumn{1}{l|}{}  & \cite{roy2019rf} & 2019 & \begin{tabular}[c]{@{}P{6.1cm}@{}}Design ConvLSTM to exploit the temporal and spatial properties of I/Q fingerprints, and compare its performance with LSTM and GRU.\end{tabular} \\ \hline 
 
\multirow{5}{*}[-10ex]{\begin{tabular}[c]{@{}A{1.5cm}@{}} Attention \end{tabular}}                                                         & \multirow{3}{*}[-10ex]{\begin{tabular}[c]{@{}A{2cm}@{}} Attention Partly\end{tabular}}              & \cite{peng2021specific} & 2021 & \begin{tabular}[c]{@{}P{6.1cm}@{}}Propose SeNet to extract characteristics in frequency domain, and further verify its performance on 20 WiFi devices.\end{tabular}                        \\ \cline{3-5}                                         &                                               \multicolumn{1}{l|}{}                                             & \cite{weng2020message} & 2020 & \begin{tabular}[c]{@{}P{6.1cm}@{}}Pay attention to the message structure and provide MSCAN, which can leverage rough prior information to enhance identification performance.\end{tabular}                                                                                                 \\ \cline{3-5} 
 &                                               \multicolumn{1}{l|}{}                                             & \cite{zhang2022data} & 2023 & \begin{tabular}[c]{@{}P{6.1cm}@{}}Propose a data-and-knowledge dual-driven architecture to exploit semantic information, and compare its performance with VGG, CNN \cite{riyaz2018deep}, and AttMsCn.\end{tabular}\\ \cline{2-5} & \multirow{2}{*}{\begin{tabular}[c]{@{}A{2cm}@{}} Transformer\end{tabular}} & \cite{shen2021radio114} & 2021 & \begin{tabular}[c]{@{}P{6.1cm}@{}}Provide a transformer-based method to identify LoRa devices with the signals of variable length, and further propose a multi-packet inference method to greatly enhance the accuracy.\end{tabular}                                                        \\ \cline{3-5}                                            &      \multicolumn{1}{l|}{}                                               & \cite{xu2021transformer} & 2021 & \begin{tabular}[c]{@{}P{6.1cm}@{}}Combine transformer and ICS algorithm to handle devices that appear outside the closed set, and verify its superiority in accuracy and robustness than Hybrid OvA \cite{hanna2020open}, Modified ICS \cite{xu2020open}, MLOSR\cite{oza2019deep},  and CROSR \cite{yoshihashi2019classification}\end{tabular} \\ \hline 
 
 \multirow{11}{*}[-20ex]{\begin{tabular}[c]{@{}A{1.5cm}@{}}  Data Augmentation\end{tabular}} & \multirow{4}{*}[-5ex]{\begin{tabular}[c]{@{}A{2cm}@{}}Added Noise-based\end{tabular}}             & \cite{shen2021radio114} & 2021 &Leverage Gaussian noise to enhance accuracy, and further propose online and offline augmentation.\\ \cline{3-5} 
&\multicolumn{1}{l|}{}  & \cite{meng2022multiuser} & 2023 & Propose LPNNs that add Gaussian noises in the latent year to improve generalization.  \\ \cline{3-5} 
 &    \multicolumn{1}{l|}{}& \cite{shen2022towards} & 2022 & Utilize channel fading models and artificial noise to realize augmentation for mobile scenarios. \\ \cline{3-5}  &                                             \multicolumn{1}{l|}{}                                               & \cite{zhang2023data187} & 2023 & Add artificial noise with random power into training fingerprints to improve the channel robustness.                                                                                                                                                                                 \\ \cline{2-5}  & \multirow{7}{*}[-15ex]{\begin{tabular}[c]{@{}A{2cm}@{}} Generated Samples-based\end{tabular}}    & \cite{yu2019robust} & 2019 & Design MSCNN to recognize ZigBee devices and use slicing techniques to obtain more samples.                                                                                                                                                                                      \\ \cline{3-5}  &                                             \multicolumn{1}{l|}{}                                               & \cite{huang2019data} & 2020 & Evaluate different data augmentation methods, including rotation, flip, and Gaussian noise.                                                                                                                                                                                             \\ \cline{3-5}  &                                             \multicolumn{1}{l|}{}                                               & \cite{zhang2022data}  & 2022 & Compare different data augmentation approaches for few-shot ADS-B scenarios.  \\ \cline{3-5}   &                                             \multicolumn{1}{l|}{}                                               & \cite{xie2019data} & 2020 & \begin{tabular}[c]{@{}P{6.1cm}@{}}Design a random integration-based method and compare its performance with conventional methods and random integration-based methods.\end{tabular}                                                                                                         \\ \cline{3-5} 
&                                             \multicolumn{1}{l|}{}                                              & \cite{piva2021tags} & 2021 &  Leverage DAG methods to enhance accuracy for time-varying environments.                                                                        
\\ \cline{3-5}  &                                             \multicolumn{1}{l|}{}                                               & \cite{liu2024overcoming}  & 2024 & Propose an adversarial augmentation method to facilitate flexible and dynamic data transformations.

\\ \cline{3-5} 
 &                                            \multicolumn{1}{l|}{}                                                & \cite{liao2019multiuser} & 2020 & \begin{tabular}[c]{@{}P{6.1cm}@{}}Propose three methods to generate additional CSI fingerprints, including averaging data augmentation, exponential averaging data augmentation, and stochastic weight averaging data augmentation. \end{tabular}                                                                                       \\ \hline

\end{longtable}

\subsection{CVNN-based Multi-Device Identification}
Considering that the real-valued neural network can not directly process complex data, some researchers utilize CVNNs to handle complex signals to obtain better extraction performance.
\begin{itemize}
\item \emph{Performance under ADS-B and WiFi scenarios:} Gopalakrishnan et al. \cite{gopalakrishnan2019robust} test the proposed CVNN architecture on external database with two different radio protocols: WiFi 802.11a (5.8 GHz) and 802.11g (2.4 GHz) commercial off-the-shelf transmitters and ADS-B (1.09 GHz) narrowband signals. The results show that the CVNN model can achieve 81.66\% and 99.53\% accuracy on ADS-B and WiFi datasets, while RVNNs can only achieve 75\% and 97.89\% accuracy, respectively.
\item \emph{CVNNs Combined with Network Compression:} Wang et al. \cite{wang2021efficient} develop an efficient approach based on CVNN and network compression named SlimCVNN to achieve both high accuracy performance and low model complexity. The simulations on synthetic fingerprint datasets verify that there is almost no accuracy gap between CVNN and SlimCVNN, while SlimCVNN has 10\%~30\% model sizes of CVNN.
\item \emph{Comparison with RVNNs:} Agadakos et al. \cite{agadakos2019deep} provide two CVNN variations: Convolutional CVNN and Recurrent CVNN for modeling signals and time series, respectively. The massive experiments demonstrate the superiority of the CVNN in identification accuracy than RVNNs under different protocols and SNR environments. Chen et al. \cite{chen2022analysis} use real LoRa and WiFi I/Q datasets to provide a deeper understanding of the impact of the fingerprint representation and the architectural layers of the models. The various experimental results verify that the CVNNs consistently obtain better identification accuracy than their “equivalent” RVNNs.
\item \emph{CVNNs Combined with FFT:} Stankowicz et al. \cite{stankowicz2019complex} utilize large and real-time I/Q datasets and explore the impact of input representation, output representation, and processing of complex values on the identification accuracy. They conclude that when using CVNN as the classifier, the representation combined I/Q and FFT can obtain better performance than I/Q.
\item \emph{CVNNs Combined with BN and Dropout Layers:} Gu et al.\cite{gu2020radio} design a CVNN-based architecture, which adds batch normalization layer after every fully-connected layer and adds the Dropout layer after each layer to optimize model parameters and avoid the over-fitting issues. The simulations on five drone signal datasets verify the advantages of the proposed method than baseline models in classification accuracy, prediction time, and confusion matrix.
\item \emph{CVNNs Combined with Residual Networks:} Wang et al. \cite{wang2020radio} propose a deep complex residual network for device identification. Compared with the approach based on contour stellar (with the identification success rate of 90.4\%) and CCVNN (with the identification success rate of 94.8\%), the suggested approach can obtain the identification success rate of 99.56\% for 20 WiFi devices.
\item \emph{CVNNs Combined with Metric Learning:} Considering insufficiently labeled training fingerprints, Fu et al. \cite{fu2023semi} introduce pseudo labels and metric learning and propose the metric-adversarial training. The proposed method focuses on exploiting the discriminative semantic characteristics. The proposed objective function is based on semi-supervised metric learning and virtual adversarial training. The simulations on the open-source large-scale real-world ADS-B datasets and WiFi datasets verify the superiority of the proposed MAT method than DRCN \cite{wang2020transfer}, SSRCNN \cite{dong2021ssrcnn}, Triple-GAN \cite{gong2019generative,li2017triple}, and SlimMIM \cite{xie2022simmim}. 
\item \textit{CVNNs Combined with Attention Mechanisms:} Jiang et al. \cite{jiang2024rf} combine CVNNs and multiple attention mechanisms to enhance authentication performance for satellite communication systems. The simulations on Xingyun satellite demonstrate its superiority than traditional CNNs and CVNNs.
\end{itemize}
\begin{les}
The CVNN-based identification methods have been studied for ADS-B datasets \cite{gopalakrishnan2019robust}, WiFi datasets \cite{gopalakrishnan2019robust,chen2022analysis}, and LoRa datasets \cite{chen2022analysis} as well as have been combined with network compression \cite{wang2021efficient}, CNNs \cite{agadakos2019deep}, RNNs \cite{agadakos2019deep}, FFT \cite{stankowicz2019complex}, and residual networks \cite{wang2020radio} to obtain higher accuracy than the RVNN-based methods. In addition, the CVNN-based methods have better adaptability and expansibility to the changes of different protocols and the number and scale of individuals. However, the training of CVNN-based methods requires more fingerprint samples and more parameters. Hence, the compression of CVNNs is an important issue.
\end{les}
\subsection{GAN-based Multi-Device Identification}
GAN is a typical deep UL model and is consisted of two modules, including the generative module and discriminative module for generating fake samples and judging whether the sample is true or false, respectively. In addition, GANs can be combined with classifiers to realize classification tasks.
\begin{itemize}
\item \emph{Identification with Fewer Fingerprints:} Zhao et al. \cite{zhao2018classification} propose AC-WGANs to realize device identification with fewer fingerprint samples and higher dimensional features. The proposed UAVs detection system includes three steps: data collection, pre-processing, and classification. The simulation results show that the proposed AC-WGANs can obtain the accuracy of 95\%.
\item \emph{Performance under Noisy Environments:} Zeng et al. \cite{zeng2022adaptive} propose a pre-processing algorithm to address the issues that the identification performance is poor in dynamic interference scenarios and the models are dependent on the quality of datasets. The proposed approach can modify the synchro-squeezed wavelet transforms through energy regularization so that the pre-processing is simplified and the robustness of fingerprints is improved. In addition, the unsupervised neural network noise feature extracting GAN is presented to obtain precise clean fingerprint characteristics from noisy signals. The simulation results show that the proposed architecture can obtain the accuracy of 96\%, 85\%, and 25\% with the SNR of 10 dB, 0dB, and -20 dB, respectively.
\item \emph{Comparison with CNNs:} Lin et al. \cite{lin2020contour} use the SSGANs \cite{tu2018semi} to achieve semi-supervised signal recognition for contour stellar images. The proposed SSGANs can exploit the inherent features of real labeled fingerprints, real unlabeled fingerprints, and fake unlabeled fingerprints. The simulation results indicate that the proposed SSGANs can obtain higher identification accuracy than CNNs when using a small number of labeled fingerprint samples.
\item \emph{Detection of Malicious Fingerprints and Identification of Multiple Legal Devices Simultaneously:} Roy et al. \cite{roy2019detection} use GANs to recognize adversarial fingerprints and identify wireless devices. The proposed generative module can generate fake fingerprints and the proposed discriminative module can distinguish the real fingerprints from the fake ones. The CNN module can achieve the classification of legitimate devices. The simulations on multiple USRP b210s verify the feasibility of the proposed architecture. Roy et al. \cite{roy2019rfal} provide an RF adversarial learning architecture to detect rogue devices and identify trusted transmitters. The designed discriminator can detect rogue devices with the accuracy of 99.9\%. The designed RNN model can reach the accuracy of 97\% for legal transmitters.
\item \emph{GANs for Unsupervised Identification:} To realize unsupervised identification, Gong et al. \cite{gong2020unsupervised} propose an unsupervised identification architecture based on InfoGANs and RF fingerprint embedding. The gray histogram is constructed from the bispectrum and then embedded into the proposed system. The priori statistical features of radio channels are leveraged in the form of the structured multi-modal latent vectors. The simulations verify the superiority of the proposed InfoGANs in evaluation score and identification accuracy.
\end{itemize}
\subsection{AE-based Multi-Device Identification}
Similar with GANs, AEs are also deep UL models. AEs are composed of encoders and decoders for dimension reduction and generation of samples, respectively. Variational Autoencoder (VAE) is a probability distribution-based AE method and has better generation performance. Besides, contractive AEs and regularized AEs are typical discriminative models.
\begin{itemize}
\item \emph{Comparison with CNNs:} Yu et al. \cite{yu2019radio} propose a general Denoising AE (DAE)-based framework for the identification of devices and further design a partially stacking approach to combine the semi-steady and steady-state fingerprints of ZegBee transmitters. The proposed partially stacking-based convolutional DAE can realize the reconstruction of the high-SNR signals and identification. The simulation results show that the proposed PAS-DAE can enhance the classification accuracy by 14\% to 23.5\% than CNN under the low SNR (from -10 dB to 5 dB) environments.
\item \emph{AEs Combined with Device Authentication Code:} Bassey et al. \cite{bassey2020device} propose an AE-based architecture for intrusion detection and present the concept of device authentication code. The reconstruction error is defined as the authentication code of devices, and Kolmogorov-Smirnov test is utilized to judge whether the fingerprints are legitimate or not. The simulations on 6 ZigBee and 5 USRP devices verify the effectiveness and robustness in channel conditions, mobility, and varying signal strength.
\item \emph{AEs for Unsupervised Identification:} Huang et al. \cite{huang2022deep} suggest the masked AE-based unsupervised pre-training approach to achieve identification with limited training fingerprints. The pre-trained network is used to predict the masked segment of fingerprints, which will be fined tuned with the supervised labels. The simulations on both synthetic fingerprints and real-world fingerprints verify the superiority of the proposed pre-training with block-wise channel-aligned masking.
\item \emph{Comparison with GANs:} Xie et al. \cite{xie2022few} propose a few-shot unsupervised identification method, where the AE module is used to obtain the latent vector of Hilbert time-frequency spectrum. The latent space can represent the hidden characteristics of fingerprints. Then, the clustering algorithm can cluster and label the latent vector, and the meta-learning method is used to classify fingerprints. The simulations under AWGAN, Rayleigh, and Rice channels demonstrate the superiority of the proposed architecture than InfoGANs \cite{gong2020unsupervised}.
\end{itemize}
\begin{les}
AEs and GANs are typical generative DL models and are usually used for dimension reduction and generation of fingerprint. AEs and GANs have been studied for UAVs \cite{zhao2018classification}, low SNR environments \cite{zeng2022adaptive}, ZigBee devices \cite{yu2019radio} and USRP devices \cite{bassey2020device} as well as have been combined with other techniques to realize identification with fewer fingerprints \cite{zhao2018classification}, unsupervised identification \cite{gong2020unsupervised,huang2022deep}, and identification and detection simultaneously \cite{roy2019detection,roy2019rfal}. AEs and GANs have outstanding unsupervised feature extraction capability and have great potential in multi-device identification.
\end{les}

Here, we summarize the contributions of all multi-device identification schemes based on CVNN, GAN, and AE in Tab. \ref{Multi-Device Identifications Schemes based on CVNN, GAN, and AE}.

\begin{longtable}{|A{1.5cm}|A{1cm}|A{1cm}|P{8.1cm}|}
\caption{Multi-Device Identifications Schemes based on CVNN, GAN, and AE} \label{Multi-Device Identifications Schemes based on CVNN, GAN, and AE} \\
\hline
\textbf{Models} & \textbf{Ref.} & \textbf{Year} & \textbf{\begin{tabular}[c]{@{}c@{}}Major Contribution\end{tabular}} \\ \hline
\endfirsthead 
\hline
\textbf{Models} & \textbf{Ref.} & \textbf{Year} & \textbf{\begin{tabular}[c]{@{}c@{}}Major Contribution\end{tabular}} \\
\hline
\endhead

\hline
\endfoot

\hline
\endlastfoot

\hline
\multirow{8}{*}[-20ex]{\begin{tabular}[c]{@{}A{1.5cm}@{}}CVNN\end{tabular}} & \cite{gopalakrishnan2019robust} & 2019 & Compare the performance between RVNN and CVNN on WiFi 802.11a (5.8 GHz) and 802.11g (2.4 GHz) protocols. \\
\cline{2-4} & \cite{wang2021efficient} & 2021 & Develop SlimCVNN based on CVNN and network compression to realize high accuracy and low complexity.  \\\cline{2-4}& \cite{agadakos2019deep} & 2019 & Provide CCVNN and RCVNN respectively for modeling signals and time series.   \\\cline{2-4} & \cite{chen2022analysis} & 2022 & Use LoRa and WiFi I/Q datasets to provide a deeper understanding of the impact of fingerprint representation and architectural layers of models.      \\ \cline{2-4} & \cite{stankowicz2019complex} & 2019 & Explore the impact of input representation, output representation, and processing of complex values on accuracy, and further conclude that the representation combined I/Q and FFT can obtain better performance than I/Q for CVNNs. \\\cline{2-4} & \cite{gu2020radio} & 2020 & Design a CVNN-based architecture, which adds BN layer after every fully-connected layer and adds the Dropout layer after each layer to optimize model parameters and avoid the over-fitting issues.  \\ \cline{2-4} & \cite{wang2020radio} & 2020 & Present a deep complex residual network, and compare its performance with CCVNN on 20 WiFi devices.                              \\\cline{2-4}
                      & \cite{fu2023semi} & 2023 & Introduce pseudo labels and metric learning to propose MAT, which focuses on exploiting the discriminative semantic characteristics, and further demonstrate its performance by the comparison with DRCN \cite{wang2020transfer}, SSRCNN \cite{dong2021ssrcnn}, Triple-GAN \cite{gong2019generative,li2017triple}, and SlimMIM \cite{xie2022simmim}. 
                      \\
\hline                      
\multirow{6}{*}{\begin{tabular}[c]{@{}A{1.5cm}@{}}  GAN\end{tabular}}  & \cite{roy2019detection} & 2019 & Use GANs to recognize adversarial fingerprints and identify devices, and verify its feasibility on multiple USRP b210s. \\
\cline{2-4}
                      & \cite{zhao2018classification} & 2018 & Propose AC-WGANs to achieve identification with fewer fingerprint samples and higher dimensional features for UAVs.                                                                                                                                                                         \\
                      
                      \cline{2-4}
                      & \cite{zeng2022adaptive} & 2022 &Design a pre-processing algorithm to tackle the issues that the authentication accuracy is poor in dynamic interference scenarios and the models are dependent on the quality of datasets, and further present NEGAN to obtain precise clean fingerprint features from noisy signals.       \\
                      
                      \cline{2-4}
                      & \cite{lin2020contour} & 2021 & Use SSGANs to realize semi-supervised signal recognition for contour stellar images.                                                                                              \\
                      
                      \cline{2-4}
                      & \cite{roy2019rfal} & 2020 & Provide RFAL to detect rogue devices and identify trusted transmitters.                                                                                                          \\
                      \cline{2-4}
                      
                      & \cite{gong2020unsupervised} & 2020 & Combine InfoGANs and RFFE to realize unsupervised identification.                                                                                                                \\
\hline 
\multirow{4}{*}[-9ex]{\begin{tabular}[c]{@{}A{1.5cm}@{}} AE\end{tabular}}  & \cite{yu2019radio} & 2019 & Present DAE and PSC-DAE realize the reconstruction of the high-SNR signals and identification.\\

                      \cline{2-4}
                      & \cite{bassey2020device} & 2021 & Propose an AE-based architecture for intrusion detection and present the concept of device authentication code. \\
                      \cline{2-4}
                      & \cite{huang2022deep} & 2022 & Design an unsupervised pre-training method based on MAE to realize identification with limited training fingerprints.                                                                                                                                                                      \\
                      \cline{2-4}
                      & \cite{xie2022few} & 2022 & Present a few-shot unsupervised identification method, where the AE module and clustering algorithm are used for dimension reduction and signal classification, respectively, and further compare its performance with InfoGANs \cite{gong2020unsupervised} under AWGAN, Rayleigh, and Rice channels.\\
\hline
\end{longtable}

\section{ML-based PLA for Attack Detection}
\label{Section IV}
In this section, we present the ML-based methods for attack detection. We divide the ML techniques into three sub-categories and compare them in detail. The organization of this section is illustrated in Fig. \ref{Organization of Section IV}.
\subsection{SL-based Attack Detection}

\begin{figure*}[h]
\centering
\includegraphics[width=0.95\textwidth]{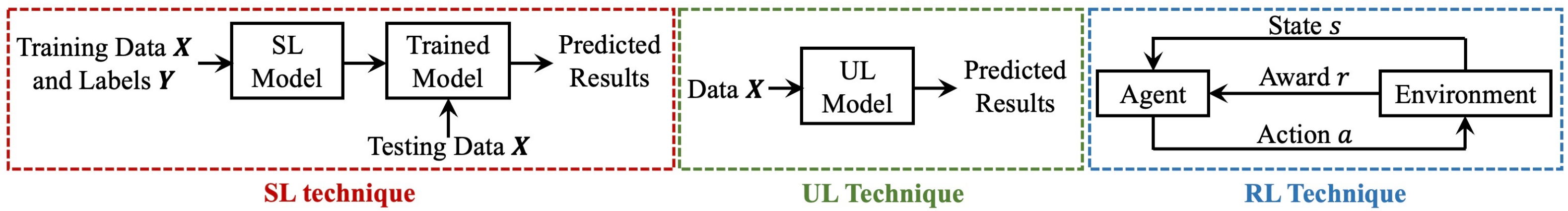}
\caption{Illustration of SL, UL, and RL techniques.}
\label{Illustration of SL, UL, and RL techniques}
\end{figure*}

As illustrated in Fig. \ref{Illustration of SL, UL, and RL techniques}, SL techniques can learn the distribution from labeled datasets. Compared with the non-ML-based attack detection methods, the ML-based approaches can automatically learn the inherent characteristics of fingerprints given their labels and obtain the optimal threshold. Here, we classify SL techniques into logistic regression, LDA, DT, KNN, SVM, ensemble learning, ELM, FL, and DL as follows.
\subsubsection{Logistic Regression}
Logistic regression is a generalized linear model. Xiao et al. \cite{xiao2017phy} provide a multi-landmark-based authentication scheme, where each landmark with multi-antenna can collect the distributed RSSIs to obtain more distinguishable characteristics. The Frank-Wolfe algorithm based on logistic regression is used to address the convex optimization problem introduced by the ML-based coefficient estimation of the regression. A distributed Frank-Wolfe-based scheme and an incremental aggregated gradient-based scheme are proposed to reduce the computation overhead, and their convergence performance and communication overheads are analyzed. The simulations on USRPs verify that the proposed distributed Frank-Wolfe-based scheme can realize the same detection performance as the Frank-Wolfe-based scheme, and the incremental aggregated gradient-based scheme can obtain a better detection performance.
\subsubsection{LDA}
LDA can map high-dimensional samples into the optimal discriminative vector space to extract classification information and compress the dimension of feature space.
\begin{itemize}
\item \emph{LDA for EI Scenarios:} Wang et al. \cite{wang2021safeguarding} propose a cluster head safeguarding mechanism realized by edge intelligence (EI) for UAVs swarm travels. The LDA algorithms is used to fuse the detection decision precisely through projecting the high-dimensional features into the low-dimensional space. Such design can keep the necessary characteristics and realize maximum separability. The simulations on UAVs verify that the proposed scheme can achieve higher detection accuracy than the baseline schemes proposed by \cite{hao2013enhanced} and \cite{zhang2020fast}.
\item \emph{LFDA based on Multi-Attributes:} By exploiting time-of-arrivals, RSS, and cyclic-features of channels, Pei et al. \cite{pei2014channel} propose two ML-based PLA methods, including linear Fisher discriminant analysis and SVM. The simulation results show that the proposed ML-based schemes can obtain lower miss detection probability and false alarm probability than the baseline scheme proposed by \cite{xiao2008using}. In addition, the proposed linear Fisher discriminant analysis-based scheme has low time complexity and space complexity.
\end{itemize}
\subsubsection{DT}
DT is a tree-based architecture describing the classification of samples. DT is composed of different nodes and directed edges, and the nodes include father node, internal nodes, and leaf nodes, denoting samples, features, and categories, respectively.
\begin{itemize}
\item \emph{Comparison with DT, SVM, and KNN:} Enad et al. \cite{enad2020machine} propose three ML-based PLA schemes, including DT, SVM, and KNN. The simulations on the OFDM systems show that all the ML-based schemes can obtain higher detection accuracy than the statistical-based scheme.
\item \emph{DTs Combined with Bootstrap Aggregating:} Du et al. \cite{du2023physical} provide a PLA scheme based on bootstrap aggregating and DT for dynamic industrial scenarios. The authentication architecture includes feature extraction, classification based on positive-unlabeled learning bagging strategy, and distinction phase. Multi-dimensional fingerprints are extracted, including amplitude, phase, CFO, and variance. The simulations on real industrial datasets verify the feasibility of the proposed strategy.
\end{itemize}
\subsubsection{KNN}
The core concept of KNN is that if most of k nearest samples of a samples in the feature space belong to a certain category, the sample also belongs to this category. Senigagliesi et al. \cite{senigagliesi2022authentication} study the authentication problem when there are relay nodes between Alice and Bob. Four cases are considered, including passive relays, autonomous active relays, coordinated active relays, and omniscient relays. Besides statistical methods, the on-class nearest neighbor-based scheme is proposed to achieve low complexity with a small number of training fingerprint samples.
\subsubsection{SVM}
SVM is a generalized linear classifier and its decision boundary is the maximum-margin hyperplane. SVM can utilize hinge loss to calculate empirical risk. In addition, SVM can realize nonlinear classification by kernel methods.
\begin{itemize}
\item \emph{Authentication under Time-Varying Channels:} Fang et al. \cite{fang2018learning} provide a ML-based adaptive PLA scheme to learn and leverage the complex time-varying environments so that the robustness and reliability of PLA are improved. The PLA problem is modeled as a linear system, and the multi-dimensional fingerprints is reduced to a single-dimensional scalar data to reduce the complexity. The adaptive algorithm based on kernel least mean square is further proposed to formulate the learning objective as a convex problem and learn the variations of fingerprints. The simulations verify that the proposed approach can greatly enhance the authentication performance during time-varying channels.
\item \emph{Authentication of Mobile Users:} Liu et al. \cite{liu2017authenticating} propose a CSI-based PLA architecture for stationary and mobile terminals. The identification includes the attack-resilient user profile builder and profile matching authenticator, which is based on SVM to realize packet level user identification. The evaluation results show that the proposed scheme is highly effective.
\item \emph{Defense Against Both Sybil and Clone Attacks:} By combining with the edge devices, Chen et al. \cite{chen2020automated} design a lightweight SVM-based PLA scheme to detect both Sybil and clone attacks in industrial wireless edge networks. The training is offline and the authentication decision is online. The simulations on two real industrial scenarios verify that the proposed scheme can realize the accuracy of 84\%.
\item \emph{Authentication under mmWave MIMO Channels:} By leveraging mmWave MIMO channels and CFO, Liu et al. \cite{liu2022online} present an online PLA framework. The Gaussian kernel approach and representer theorem is utilized to model the PLA problem as a binary classification model. An online algorithm is further developed based on the constructed convex objective function. The analytical expressions for false alarm and detection rates are also derived. The simulations verify that the proposed solution can defend against spoofing attacks in mmWave MIMO scenarios.
\item \emph{Fuzzy Learning:} Fang et al. \cite{fang2020fuzzy} provide a fuzzy learning-based PLA scheme to combine muti-dimensional fingerprints. The designed model is based on imperfect feature estimation to make decisions, thus providing a higher-level protection for legal devices. A hybrid learning method is further provided to update parameters of systems, where the least-square estimator and gradient descent are used for linear and nonlinear parameters, respectively.
\end{itemize}
\subsubsection{Ensemble Learning}
The core concept of ensemble learning is to construct and combine multi-learners. Specifically, a group of “individual learners” are generated first, and then they are combined through a sort of strategies, such as average method and voting method.
\begin{itemize}
\item \emph{Threshold-Free Authentication:} Pan et al. \cite{pan2019threshold} propose four threshold-free PLA schemes with the help of ML techniques, including DT, SVM, KNN, and ensemble learning. The authentication performance of the PLA based on channel matrices and channel differences are also compared. The channel differences are represented as (7), (8), and (9). The simulations verify that the proposed authentication methods can significantly improve the authentication accuracy and are suitable for wireless industrial cyber-physical systems. 
    \begin{equation}
        \bm{H}_{\mathrm{diff}}^{1}=\frac{||\bm{H}_{t}-\bm{H}_{1}||_{2}}{||\bm{H}_{1}||_{2}}
    \end{equation}
    \begin{equation}
\begin{cases}\boldsymbol{\bm{H}}_{\mathrm{diff}}^2=\left|\frac{\sum_{m,n}\lvert\boldsymbol{\bm{H}}_t(m,n)-\boldsymbol{\bm{H}}_{t-1}(m,n)e^{j\boldsymbol{\phi}(t)}\rvert}{\sum_{m,n}\lvert\boldsymbol{\bm{H}}_{t-1}(m,n)-\boldsymbol{\bm{H}}_{t-2}(m,n)e^{j\boldsymbol{\phi}(t-1)}\rvert}-1\right|\\\phi(t)=\arg\bigl(\boldsymbol{\bm{H}}_t(m,n),\boldsymbol{\bm{H}}_{t-1}^*(m,n)\bigr)\end{cases}
    \end{equation}
    \begin{equation}
        \bm{H}_{\mathrm{diff}}^{3}=\left|\frac{\sum_{m,n}|\bm{H}_{t}(m,n)-\bm{H}_{t-1}(m,n)|^{2}}{\sum_{m,n}|\bm{H}_{t-1}(m,n)-\bm{H}_{t-2}(m,n)|^{2}}-1\right|
    \end{equation}
\item \emph{Authentication based on WV under Industrial Scenarios:} Xie et al. \cite{xie2021weighted} proposed a weighted voting (WV)-based PLA approach for edge computing scenarios. The proposed WV method based on simple lines and simple parts can realize cooperative decision, thus improving the accuracy and reducing the complexity. The simulations on public datasets and field-measured datasets verify that the proposed scheme can realize better performance than the baseline schemes, including direct classification, averaged classification, and down-sampled classification, proposed by \cite{germain2020multi,valdovinos2009combining,lau2013weighted}, respectively.
\item \textit{Cooperative Authentication based on WV:} Zhou et al. \cite{zhou2024securing} propose a decision-level-based cooperative PLA scheme aimed at enhancing secure access. A dynamic WV mechanism is introduced to address challenges posed by unreliable cooperators. This approach effectively manages cooperators, ensuring concise and efficient cooperation management to bolster system security.

\end{itemize}
\subsubsection{ELM}
ELM is a special FCNN, and it has a single latent layer. The weight parameters in the latent layer do not need to be updated, and the learning process only calculates the output weights. Wang et al. \cite{wang2017physical} present an ELM-based PLA architecture to exploit the multi-dimensional features of channel environments, thus detecting the spoofing attackers in dynamic networks. The simulations show that the proposed ELM-based scheme can obtain lower minimum Bayes risk than the baseline scheme proposed by \cite{xiao2016phy}.
\subsubsection{FL}
FL is a distributed ML technique, where multi-devices can realize the joint training models by interacting the intermediate parameters without sharing training datasets. Wang et al. \cite{wang2021collaborative} provide a horizontal FL-based collaborative PLA scheme to release computational pressure on IoT devices with limited computation and storage resources. The PLA problem is formulated as the training problem of the classifier, where the aim is to obtain the weight parameters. Then, a distributed identification architecture is proposed to assign the classification task to trusted collaborators. Finally, the whole local parameters are aggregated at the center device. The simulation results show that the proposed horizontal-FL-based scheme can obtain the miss detection rate and false alarm rate of less than 1\%.
\subsubsection{DL}
The introductions of DL have been presented in Section III. DNNs can extract high-level characteristics of fingerprints to realize attack detection.
\begin{itemize}
\item \emph{FCNN, CNN, and CPNN for Industrial Scenarios:} Liao et al. \cite{liao2019deep} propose a DL-based PLA framework to guarantee the identity security of industrial wireless sensor networks, and further provide three PLA algorithms based on DL, including FCNN, CNN and convolution pre-processing neural network (CPNN). The proposed CPNN-based PLA requires few computational resources. The mini batch trick is utilized to improve the training speed. The simulation on USRPs demonstrates the effectiveness of the proposed PLA algorithms.
\item \emph{Comparison with KNN and SVM:} Pan et al. \cite{pan2020physical} propose a residual network-based PLA scheme to identify mobile devices in industrial wireless CPS. The simulations on real industrial datasets demonstrated its superiority than KNN and SVM.
\item \emph{Authentication for Industrial Scenarios:} Pan et al. \cite{pan2020clone} present a clone detection scheme based on back propagation neural network and physical-layer reputation for industrial wireless CPS. The physical-layer reputation is accumulated by CSI, and the attack detection is performed by group detection. The numerical simulations on USRPs verify that the proposed scheme can greatly enhance the detection accuracy.
\item \emph{Authentication based on Confidence Score Branch:} Wang et al. \cite{wang2022csi} design an FCNN-based PLA scheme to realize fast and lightweight identification. The CSI information is mapped to the location of devices, and is further mapped to its identity. The mapping relationship between the CSI fingerprints and the identity is learned by the proposed authenticator with a confidence score branch. The simulation results indicate that the proposed scheme is robust to channel estimation errors.
\item \emph{Authentication based on SNR Trace:} Wang et al. \cite{wang2021exploiting} propose a PLA scheme based on the SNR trace obtained at the receiver in the sector level sweep process, and further present two ML-aided approaches. For the first approach, a novel DNN framework is proposed, including back propagation neural network, forward propagation networks, and GANs. For the second approach, the Siamese network is provided to address the issue that historical fingerprints can not support the authentication in a new communication session. The simulations under different scenarios verify that the proposed scheme can realize the detection accuracy of 99\%.
\item \textit{Authentication based on Semantic Fingerprints:} Gao et al. \cite{gao2024esanet} employ a single-stage object detection network to extract the semantic knowledge of CSI in MIMO systems. The suggest approach substantially minimizes data processing overhead and authentication latency.
\item \textit{Authentication of Mobile Devices:} Wang et al. \cite{wang2024spatiotemporal} extract correlation and scattering characteristics of mobile devices and convert them to a CSI sequence, which is further used for classification by CNNs. Jing et al. \cite{jing2024multi} further present a ResNet-based PLA model to authenticate multiple mobile transmitters. The training mechanism not only improves the accuracy of authentication but also accelerates the convergence speed of ResNet. Han et al. \cite{han2024model} introduce a model-driven learning algorithm, which focuses on extracting pertinent channel features to alleviate inter-symbol and inter-carrier interferences. Experimental findings demonstrate that the proposed scheme outperforms existing data-driven models across different SNRs and velocities.
\end{itemize}

Here, we summarize the contributions of all SL-based attack detection schemes in Tab. \ref{The SL-based Attack Detection Schemes}.

\begin{longtable}{|A{1.5cm}|A{1cm}|A{1cm}|P{8.1cm}|}
\caption{The SL-based Attack Detection Schemes} \label{The SL-based Attack Detection Schemes} \\
\hline
\textbf{Models} & \textbf{Ref.} & \textbf{Year} & \textbf{\begin{tabular}[c]{@{}P{8.1cm}@{}}Major Contribution\end{tabular}} \\ \hline
\endfirsthead 
\hline
\textbf{Models} & \textbf{Ref.} & \textbf{Year} & \textbf{\begin{tabular}[c]{@{}P{8.1cm}@{}}Major Contribution\end{tabular}} \\
\hline
\endhead

\hline
\endfoot

\hline
\endlastfoot

\hline

Logistic Regression                & \cite{xiao2017phy} & 2018 & Propose a multi-landmark-based authentication scheme and introduce FW and dFW algorithms based on logistic regression to address the convex optimization problem and reduce the communication overhead, respectively.                                                                 \\
\hline
\multirow{2}{*}[-3ex]{\begin{tabular}[c]{@{}A{1.5cm}@{}}LDA\end{tabular}}               & \cite{enad2020machine} & 2021 & Present a CF safeguarding mechanism achieved by EI for UAVs swarm travels, and verify its superiority than the baseline schemes proposed by \cite{hao2013enhanced,zhang2020fast}.                                                                                                                 \\
\cline{2-4}
                                   & \cite{pei2014channel} & 2014 & Design two attack detection schemes based on LFDA and SVM by exploiting time-of-arrivals, RSS, and cyclicfeatures.                          \\
\hline
\multirow{2}{*}[-2ex]{DT}                & \cite{enad2020machine} & 2020 & Compare three ML-based PLA schemes for OFDM systems, including DT, SVM, and KNN.                                                                                                                                                                                                       \\
\cline{2-4}
                                   & \cite{du2023physical} & 2023 & Combine bootstrap aggregating and DT to detect signals for dynamic industrial scenarios.                                                                                                                                                                                               \\
                                   \hline
                                  
KNN                                & \cite{senigagliesi2022authentication} & 2022 & Study the authentication problem by considering relay nodes between Alice and Bob and four cases, including passive relays, autonomous active relays, coordinated active relays, and omniscient relays.                                                                               \\
\hline
\multirow{5}{*}[-10ex]{\begin{tabular}[c]{@{}A{1.5cm}@{}} SVM\end{tabular}}            & \cite{fang2018learning}   & 2019 & Provide a kernel-based adaptive PLA scheme for complex time-varying environments.                                                                                                                                                                                                      \\
\cline{2-4}
                                   & \cite{liu2017authenticating} & 2018 & Propose an authentication architecture based on CSI and SVM for stationary and mobile terminals.                                                                                                                                                                                       \\
                                   \cline{2-4}
                                   & \cite{chen2020automated} & 2021 & Show an SVM-based PLA framework to detect both Sybil and clone attacks in industrial wireless edge networks.                                                                                                                                                                      \\
                                   \cline{2-4}
                                   & \cite{liu2022online} & 2022 & Develop an online PLA architecture based on Gaussian kernel and CFO for mmWave MIMO channels.                                                                                                                                                                                          \\
                                   \cline{2-4}
                                   & \cite{fang2020fuzzy} & 2020 & Provide a fuzzy learning-based approach for multi-dimensional fingerprints, and further provide a hybrid learning method to update parameters of systems.                                                                                                                              \\
                                   \hline
\multirow{2}{*}[-6ex]{\begin{tabular}[c]{@{}A{1.5cm}@{}}Ensemble Learning\end{tabular}} & \cite{pan2019threshold}  & 2019 & Propose four threshold-free PLA schemes based on ML techniques, including DT, SVM, KNN, and random forest.                                                                                                                                                                           \\
\cline{2-4}
                                   & \cite{xie2021weighted} & 2022 & Present a WV-based  approach for edge computing scenarios, and further confirm its effectiveness on public datasets and fiddle-measured datasets by the comparison with direct classification \cite{germain2020multi}, averaged classification \cite{valdovinos2009combining},  and down-sampled classification \cite{lau2013weighted}. \\
                                   \hline
ELM                                & \cite{wang2017physical} & 2017 & Introduce an ELM-based PLA scheme to defend against spoofing attackers in dynamic networks, and further demonstrate that it can obtain lower minimum Bayes risk than the baseline scheme proposed by \cite{xiao2016phy}.                                                                         \\
\hline
FL                                 & \cite{wang2021collaborative} & 2021 & Develop an HFL-based collaborative PLA scheme to release computational pressure on IoT devices.                                                                   
\\
\hline

\multirow{9}{*}{\begin{tabular}[c]{@{}A{1.5cm}@{}} DL\end{tabular}}      & \cite{liao2019deep} & 2019 & Provide a DL-based PLA framework, and further propose three PLA. algorithms based on FCNN, CNN, and CPNN.                                                                                                                                                                             \\
\cline{2-4}
                                  
                                   & \cite{pan2020physical} & 2020 &Propose a residual network-based method for the authentication of mobile devices in industrial wireless CPS, and further verify its superiority in attack detection than KNN and SVM on real industrial datasets.\\

                                   \cline{2-4}
                                   & \cite{pan2020clone} & 2021 & Present a clone detection scheme based on BPNN and physical-layer reputation.\\

                                   \cline{2-4}
                                   & \cite{wang2022csi} & 2022 &Design an FCNN-based detection approach, where the mapping relationship between CSI fingerprints and the corresponding identity is learned by the confidence score branch.  \\
                            
\cline{2-4}
& \cite{wang2021exploiting} & 2021 &Propose a PLA architecture based on SNR traces, and further present an authentication algorithm based on BPNN, forward propagation networks, and GANs.\\                                            

\cline{2-4}
& \cite{gao2024esanet} & 2024 & Employ a single-stage object detection network to extract the semantic knowledge of CSI in MIMO systems.\\   

\cline{2-4}
& \cite{wang2024spatiotemporal} & 2024 & Extract correlation and scattering characteristics of mobile multi-users and convert them to CSI sequences. \\  

\cline{2-4}
& \cite{jing2024multi} & 2024 & Propose the training mechanism to improve the accuracy of authentication and accelerate the convergence speed of ResNet. \\ 

\cline{2-4}
& \cite{han2024model} & 2024 & Introduce a model-driven learning algorithm, which focuses on extracting pertinent channel features to alleviate inter-symbol and inter-carrier interferences. \\ 

\hline
\end{longtable}

\begin{les}
The SL-based attack detection schemes have been studied for UAVs \cite{wang2021safeguarding}, OFDM systems \cite{enad2020machine}, IIoT \cite{xie2021weighted,pan2019threshold}, time-varying environments \cite{fang2018learning}, mobile users \cite{liu2017authenticating}, mmWave MIMO channels \cite{liu2022online}, 5G links \cite{abdrabou2022adaptive}, and EI scenarios \cite{xie2021weighted}. Among the SL-based methods, SVM and DNNs are most widely used models due to the efficient nonlinear learning ability. However, the SL algorithms usually require many labels of transmitters and attackers, limiting the actual application.
\end{les}
\subsection{UL-based Attack Detection}
In wireless systems, it is reasonable to assume that the prior information of attackers is unknown or the fingerprints of attackers are unlabeled. For this reason, SL algorithms are not always feasible due to they require labeled datasets. To address this issue, UL algorithms are promising methods. The related works are as follows.
\subsubsection{Clustering}
Clustering is a typical UL classification algorithm, and it can divide fingerprints into several disjoint subsets (also called clusters) to realize malicious node detection.
\begin{itemize}
\item \emph{Clustering of RSSs:} Yang et al. \cite{yang2012detection} propose a PLA scheme based on RSS and clustering to determine the number of attackers for wireless networks with multiple spoofing attackers. The number of clusters is estimated through partition energy and merging energy \cite{kaijun2007estimating}. The simulations under two real office buildings with WiFi networks and ZigBee networks verify that the proposed method can obtain the hit rate and precision of over 90 percent.
\item \emph{Clustering of Multi-Attributes:} Xia et al. \cite{xia2021multiple} study the correlation of multi-fingerprints and propose an unsupervised PLA scheme based on non-parametric clustering algorithm. Based on the evolution algorithm provided in \cite{wang2009estimating}, the improved system evolution method is proposed to reduce the complexity. The simulations on synthetic datasets and real industrial datasets show that the proposed scheme can realize the $F_1$ measure of more than 99\% without the prior information of attackers.
\item \emph{Clustering Algorithms Combined with RNNs:} Wang et al. \cite{wang2018supervised} combine DNN and K-means to realize semi-supervised learning-based PLA. The convolutional recurrent neural network is utilized to learn the local characteristics and the dependencies between different frequencies in CSI fingerprints. The simulation results reveal that the proposed scheme can yield excellent detection performance with limited labeled fingerprints.
\end{itemize}
\subsubsection{OCC-SVM}
Given only Alice’s fingerprint samples, OCC-SVM algorithms can identify data points distinct from the legal fingerprint samples by constructing a boundary in high-dimensional feature space, allowing it to perform well in attack detection when Eve’s fingerprint samples are scarce.
\begin{itemize}
\item \emph{Authentication under LEO Satellite Scenarios:} Abdrabou et al. \cite{abdrabou2022leo} propose an effective PLA solution based on single-class classification SVM to defend against spoofing attacks in low-earth orbit (LEO) satellite scenarios. The proposed SCC-SVM scheme can learn the inherent features of RSS and Doppler frequency spread, and the performance for on-the-pause satellite communication systems is further evaluated. The simulation results reveal that utilizing both RSS and Doppler frequency spread can obtain lower missed detection rate and false alarm rate.
\item \emph{Performance versus Different Kernel Functions:} Hoang et al. \cite{hoang2021physical} propose an architecture to convert radio signals into structured datasets and further present an SVM-based authenticator. Two classes of SVM classifiers are considered, including a classic twin-class SVM and a single-class SVM. The performance of the proposed schemes is evaluated over different choices of the kernel function, fingerprints, and the eavesdroppers’ power.
\item \emph{Comparison with Statistical Decision Methods:} Senigagliesi et al. \cite{senigagliesi2020comparison} evaluate and compare two different statistical decision approaches for PLA, and further propose two ML-based schemes based on nearest neighbor and SVM. The numerical simulations indicate that the ML-based schemes can realize the lowest probability of missed detection when there is a small spatial correlation between the main channel and the adversary one, otherwise, the statistical approaches are superior.
\item \emph{OC-SVMs Combined with Antenna Diversity Techniques:} Abdrabou et al. \cite{abdrabou2022adaptive} propose an adaptive lightweight PLA approach that exploit antenna diversity technique to enhance the recognizable fingerprints. The one-class classifier SVM is utilized for outlier and anomaly detection. The simulations on sounding reference signals in the 5G uplink radio frame show the superiority of the proposed scheme than the baseline schemes proposed by \cite{pei2014channel} and \cite{senigagliesi2020comparison}.
\end{itemize}
\subsubsection{Manifold Learning}
Manifold learning can recover the low-dimensional manifold structure from high-dimensional sampled data, and further find the corresponding embedded mapping, thus realizing dimension reduction or data visualization. Xia et al. \cite{xia2022physical} consider the time-varying characteristics of fingerprints caused by the mobility of UAVs and formulate the identification problem as the recognition of nonlinearly separable data. The manifold learning is utilized to address this issue by establishing the Markov chain of fingerprints in the time domain and evaluating the state transition probability of UAVs. The simulation results show that the proposed scheme can improve the detection accuracy by more than 18\% compared with baseline schemes.
\subsubsection{GMM}
GMM is a special clustering algorithm, which can decompose the object into several sub-models based on Gaussian probability density functions.
\begin{itemize}
\item \emph{Online Authentication:} Gulati et al. \cite{gulati2013gmm} use GMM to formulate the probabilistic model of radio channels of transmitters. The proposed scheme can realize online learning and parameter updating. The simulation results show that the proposed scheme can realize the miss detection rate of 0.1\% for the false alarm rate of 0.4\%.
\item \emph{Authentication under MC-MTC scenarios:} Weinand et al. \cite{weinand2017physical} propose a GMM-based detection method to defend against data manipulation and masquerade attacks for mission critical machine type communications in wireless networks. The simulation results show that the proposed scheme can reach detection rate of 99.98\%.
\item \emph{Comparison with RL Techniques:} To realize high robustness to the errors of channel estimation, Qiu et al. \cite{qiu2018physical} present a PLA scheme based on pre-processing channel variations and multi-dimensional fingerprints. The PLA problem is further formulated as the comparison of two-dimensional feature vectors. The proposed scheme based on probabilistic model requires only a few training fingerprints of legitimated users. The simulation results show its superiority than the schemes proposed in \cite{xiao2016phy} and \cite{weinand2017physical}.
\end{itemize}
\subsubsection{GP}
GP is the combination of a series of random variables that obey Gaussian distribution in an index set. The classic GP models include GP Regression (GPR) and GP Classification (GPC). GPR can be used to predict fingerprints, while GPC can identify fingerprints directly.
\begin{itemize}
\item \emph{GPR:} Wang et al. \cite{wang2021channel} provide a PLA scheme based on GP channel prediction for IoT devices. Specifically, historical CSI fingerprints and geographical data of devices are leveraged to construct a mapping to predict the next legal CSI fingerprint for identification. The one-class authentication scheme is further proposed, which requires no channel information of attackers. The simulations on quasideterministic radio channel generator verify the superiority of the proposed scheme than the baseline schemes proposed in \cite{liu2016physical} and \cite{tomasin2018analysis}.
\item \emph{GPC:} Qiu et al. \cite{qiu2018physical} propose a GP-based PLA scheme to track multi-targets and reduce the identification overhead. The proposed scheme can intelligently learn the dynamic time-frequency fingerprints of channels. 
\end{itemize}

\subsubsection{AE}

The introductions of AE have been provided in Section III-H. The latent space with lower dimensions can be used for authentication. Meng et al. \cite{meng2022physical} suggest a PLA scheme based on hierarchical variational autoencoder to defend against spoofing attacks in IIoT without requiring attackers’ training fingerprints. The designed architecture is a cascade neural network based on AE and VAE. The AE module serves as a low-level extractor, and the VAE module is for further dimension reduction and authentication. The VAE module is equipped with a single-peak and a revised double-peak Gaussian distribution for fingerprint reproduction and classification, respectively. The constructed loss function is derived, including an approximation and an exact calculation. The simulation results demonstrate its superiority than the ISE scheme proposed in \cite{xia2021multiple}.

Here, we summarize the contributions of all UL-based attack detection schemes in Tab. \ref{The UL-based Attack Detection Schemes}.

\newpage
\begin{longtable}{|A{1.5cm}|A{1cm}|A{1cm}|P{8.1cm}|}
\caption{The UL-based Attack Detection Schemes} \label{The UL-based Attack Detection Schemes} \\
\hline
\textbf{Models} & \textbf{Ref.} & \textbf{Year} & \textbf{\begin{tabular}[c]{@{}P{8.1cm}@{}}Major Contribution\end{tabular}} \\ \hline
\endfirsthead 
\hline
\textbf{Models} & \textbf{Ref.} & \textbf{Year} & \textbf{\begin{tabular}[c]{@{}P{8.1cm}@{}}Major Contribution\end{tabular}} \\
\hline
\endhead

\hline
\endfoot

\hline
\endlastfoot

\hline
 & \cite{xia2021multiple}  & 2021 & Study the correlation of multi-fingerprints and propose an authentication architecture based on non-parametric clustering algorithms, and propose an improved system evolution method based on the evolution algorithm \cite{wang2009estimating}.\\
\cline{2-4} Clustering & \cite{yang2012detection} & 2013 & Leverage clustering algorithms to determine the number of attackers, and further demonstrate its effectiveness under two real office buildings with WiFi networks and ZigBee networks.\\
                            \cline{2-4}&\cite{wang2018supervised} & 2018 & Combine DNN and K-means to propose CRNN to learn to local features and the dependencies between different frequencies in CSI fingerprints.\\
                            \hline
   \multirow{4}{*}[-5ex]{\begin{tabular}[c]{@{}A{1.5cm}@{}} OCC-SVM\end{tabular}}& \cite{abdrabou2022adaptive} & 2022 & Design a lightweight PLA scheme based on OCC-SVM and antenna diversity technique, and further demonstrate its advantages than the baseline schemes proposed by \cite{pei2014channel,senigagliesi2020comparison} for SRSs.  \\
\cline{2-4}& \cite{abdrabou2022leo} & 2022 & Introduce an SCC-SVM-based PLA solution for LEO satellite scenarios.  \\ \cline{2-4}& \cite{hoang2021physical} & 2021 & Consider two classes of SVM classifiers, including TC-SVM and SC-SVM, and further evaluate the performance over different choices of kernel functions, fingerprints, and the attackers’ power.  \\
                          \cline{2-4}
                            & \cite{senigagliesi2020comparison} & 2021 &Evaluate and compare two different statistical decision methods, and further propose two ML-based PLA schemes based on nearest neighbor and SVM.  \\
                           \hline

Manifold Learning  & \cite{xia2022physical} & 2022 & Consider the time-varying characteristics of fingerprints caused by the mobility of UAVs and formulate the authentication problem as the recognition of nonlinearly separable data, and further use manifold learning to establish the Markov chains to address this issue.\\
\hline
& \cite{gulati2013gmm} & 2013 & Use GMM models to formulate the probabilistic model of radio channels of transmitters and realize online learning and parameter updating.\\
\cline{2-4} 

GMM  & \cite{weinand2017physical} & 2017 & Design a GMM-based method to defend against data manipulation and masquerade attacks for MC-MTC scenarios. \\  \cline{2-4} & \cite{qiu2018physical} & 2018 & Present a PLA method based on pre-processing channel variations and multi-dimensional fingerprints to realize high robustness to the errors of channel estimation. \\
                           \hline
GP & \cite{wang2021channel} & 2022 &  Provide a PLA scheme based on GP channel prediction for IoT devices and propose an OCA scheme without knowing channel information of attackers, and further compare its performance with the schemes proposed by \cite{liu2016physical} and \cite{tomasin2018analysis}. \\
 \cline{2-4}& \cite{qiu2018physical} & 2018 & Develop a GP-based PLA scheme to track multi-targets and reduce the authentication overhead for dynamic channels. \\ \hline
AE  & \cite{meng2022physical} & 2023 & Suggest an HVAE-based scheme to defend against spoofing attacks in IIoT, and further derive the loss function consisting the AE module and the VAE module with a single-peak and a revised double-peak Gaussian distribution.  \\
\hline
\end{longtable}

\begin{les}
Compared with the SL-based schemes, the UL-based attack detection schemes do not require the prior information or training fingerprint samples of attackers, thus having more universality in actual applications. The UL-based schemes have been studied for ZigBee networks \cite{yang2012detection}, industrial networks \cite{xia2021multiple,meng2022physical}, UAVs \cite{xia2022physical}, LEO satellite scenarios \cite{abdrabou2022leo}, MC-MTC scenarios \cite{weinand2017physical}, and IoT \cite{wang2021channel}. Compared with the non-DL-aided UL models, including clustering algorithms, manifold learning, and GMM algorithms, AEs can extract features of fingerprints with higher dimensions and are more suitable for future Massive MIMO-enabled scenarios.
\end{les}
\subsection{RL-based Attack Detection}
Compared with the SL and UL techniques, RL techniques do not require accurate input and precise parameter updates. As illustrated in Fig. \ref{Illustration of SL, UL, and RL techniques}, the reward obtained by the agent through the interactions with the environments guides the behavior, and the goal is that the agent can obtain the maximum reward. We classify the RL techniques into non-DL-aided RL and DRL, and the latter combines the perception ability of DL and the decision ability of RL.
\subsubsection{Non-DL-Aided RL}
\begin{itemize}
\item \emph{Effectiveness of RL Methods:} Xiao et al. \cite{xiao2016phy} formulate the interactions between the legal receiver and spoofing attackers as a zero-sum game and propose a spoofing detection method based on Q-learning and Dyna-Q. The receiver chooses the threshold based on the Bayesian risk, while the attackers determine the spoofing frequency to minimize the utility of the legal receiver. The optimal threshold is obtained through the designed RL method. The simulations on USRPs validate the feasibility of the proposed strategies.
\item \emph{Intelligent Attacks:} Gao et al. \cite{gao2020physical} study the PLA in the threat of intelligent spoofing attacks. The optimal transmit power allocation is derived, and the optimal intelligent attack for legitimate devices is found. A cooperative PLA scheme is further proposed to defend against the above-mentioned attacks with low time complexity and high accuracy. The closed-form expression for the belief threshold is also provided. The simulation results verify the superiority of the proposed scheme than the schemes proposed by \cite{xiao2009channel,wang2017physical,yan2014optimal}.
\item \emph{Authentication under underwater acoustic sensor networks:} Li et al. \cite{li2015spoofing} formulate the interaction between the surface station and the underwater spoofers in underwater acoustic sensor networks as a zero-sum game, and further derive the Nash equilibrium. The existing condition for the unique Nash equilibrium is also presented.
\item \emph{Authentication under MIMO systems:} Xiao et al. \cite{xiao2017game} propose a Q-learning-based PLA scheme to detect spoofing attackers in MIMO systems. The optimal threshold can be obtained without knowing system parameters. The Dyna architecture and prioritized sweeping is further applied to improve the detection accuracy during time-varying channels. The simulation results show that the proposed scheme can improve detection performance compared with Q-learning algorithm.
\item \emph{Authentication under UAVs:} Zhou et al. \cite{zhou2022game} provide a PLA scheme based on RL and RSS for UAVs. The false alarm rate and miss detection rate are derived, and the Nash equilibrium and its existence condition for the designed zero-sum authentication game are further presented. The Monte Carlo simulation results demonstrate the analytical expressions.
\item \textit{Collaborator Selection:} Zhang et al. \cite{zhang2023cooperative} propose a FL-based cooperative PLA scheme, and further employ Q-learning algorithm to select the optimal collaborator. The simulation analysis and real-world communication experiments verify the superiority than blind cooperation. 
\end{itemize}
\subsubsection{DRL}
\begin{itemize}
\item \emph{Authentication under VANETs:} Lu et al. \cite{lu2020reinforcement} propose a DRL-based PLA scheme to defend against rogue edge attackers in VANETs. The channel information is shared for the mobile devices and onboard unit with the same moving trace, and RL is utilized to choose authentication modes and parameters. Transfer learning and DL are applied to improve efficiency and further improve authentication performance, respectively. The simulation and experimental results show that the proposed scheme can improve detection accuracy compared with the baseline schemes proposed in \cite{riyaz2018deep,liu2017active,lu2018learning}.
\item \emph{Authentication under IoT Scenarios:} Wu et al. \cite{wu2023game} consider three cases in static channels: including multi-player games, zero-sum games with collisions, and zero-sum games without collisions, and further derive the closed-form expressions for Nash equilibrium. A multi-agent deep deterministic policy gradient algorithm is further proposed for dynamic environments. The simulation experiments prove that the proposed PLA schemes are feasible for IoT scenarios.
\item \emph{Authentication under UWSNs:} Xiao et al. \cite{xiao2018learning} present a PLA framework to detect spoofing attackers for UWSNs. The power delay profile of underwater acoustic channels is utilized to recognize sensors, and RL is applied to choose the parameters of authentication systems. DRL is further utilized to improve the authentication accuracy. The simulation results indicate that the proposed scheme increase the utility of the system compared with the benchmark scheme proposed in \cite{li2015spoofing}.
\end{itemize}

Here, we summarize the contributions of all RL-based attack detection schemes in Tab. \ref{The RL-based Attack Detection Schemes}.

\newpage
\begin{longtable}{|A{1.5cm}|A{1cm}|A{1cm}|P{8.1cm}|}
\caption{The RL-based Attack Detection Schemes} \label{The RL-based Attack Detection Schemes} \\
\hline
\textbf{Models} & \textbf{Ref.} & \textbf{Year} & \textbf{\begin{tabular}[c]{@{}P{8.1cm}@{}}Major Contribution\end{tabular}} \\ \hline
\endfirsthead 
\hline
\textbf{Models} & \textbf{Ref.} & \textbf{Year} & \textbf{\begin{tabular}[c]{@{}P{8.1cm}@{}}Major Contribution\end{tabular}} \\
\hline
\endhead

\hline
\endfoot

\hline
\endlastfoot

\hline
\multirow{6}{*}[-15ex]{\begin{tabular}[c]{@{}A{1.5cm}@{}} Non-DL-Aided RL\end{tabular}} & \cite{xiao2016phy}  & 2016 & Formulate the interactions between the legal receiver and spoofing attackers as a zero-sum game and propose a spoofing attack method based on Q-learning and Dyna-Q.\\
\cline{2-4}& \cite{xiao2017game}  & 2017 & Propose a Q-learning-based scheme to detect spoofing attackers in MIMO systems, and further apply Dyna-PS algorithm to improve the detection accuracy during time-varying channels. \\ \cline{2-4}& \cite{gao2020physical} & 2020 & Study the PLA in the threat of intelligent spoofing attacks, and provide comprehensive theoretical deduction, including  the optimal transmit power allocation, the optimal intelligent attack, and the belief threshold.\\ \cline{2-4} & \cite{li2015spoofing} & 2015 & Suggest a Q-learning-based attack detection scheme for UWSNs, and derive the Nash equilibrium and existing condition.\\ \cline{2-4}
& \cite{zhou2022game} & 2022 & Provide a PLA scheme based on RL and RSS for UAVs, and derive the false alarm rate and miss detection rate. \\ \cline{2-4}
& \cite{zhang2023cooperative} & 2023 & Employ Q-learning algorithm to select the optimal authentication collaborator.\\  \hline
\multirow{3}{*}[-10ex]{\begin{tabular}[c]{@{}A{1.5cm}@{}}  DRL\end{tabular}}       & \cite{lu2020reinforcement} & 2020 & Present a DRL-based PLA scheme to detect rogue edge attackers in VANETs, and demonstrate its superiority in attack detection by the comparison with the baseline schemes proposed by \cite{xiao2016phy,liu2017active,lu2018learning}. \\
\cline{2-4}& \cite{wu2023game} & 2023 & Consider three cases in static channels, including multi-player games, zero-sum games with collisions, and zero-sum  games without collisions, and further propose an MADDPG-based scheme for dynamic environments. \\ \cline{2-4}
                                 & \cite{xiao2018learning} & 2019 & Develop a RL-based PLA scheme to detect spoofing attackers for UWSNs and further utilize DRL to improve the authentication accuracy, and further verify its superiority than the benchmark scheme proposed by \cite{li2015spoofing}. \\
                                \hline
    \end{longtable}
    
\begin{les}
The RL-based attack detection schemes usually formulate the game between the legitimate receiver and spoofing attackers and have been studied for USWSNs \cite{li2015spoofing,xiao2018learning}, MIMO systems \cite{xiao2017game}, UAVs \cite{zhou2022game}, VANETs \cite{lu2020reinforcement}, and IoT \cite{wu2023game} as well as have been combined with other techniques to obtain higher accuracy and efficiency, such as TL \cite{lu2020reinforcement} and DL \cite{lu2020reinforcement,wu2023game,xiao2018learning}. The intelligent spoofing attacks \cite{gao2020physical} have also been studied, including the derivations of the optimal transmit power allocation and the optimal intelligent attack for legitimate devices as well as the closed-form expression for the belief threshold.
\end{les}

\section{Fingerprint Datasets}

\label{Section V}
The open-source, high-quality, and large-scale fingerprint datasets are indispensable parts of promoting the development of the ML-based PLA. In this section, we sort out and summarize various open-source fingerprint datasets in detail, aiming at proving researchers with comprehensive knowledge for further application of the PLA.
\subsection{RF Fingerprint Datasets}
\label{Section V-A}
In earlier literature, the DL-based multi-device identification methods obtain high accuracy for small-scale devices under LOS environments. However, it is challenging to extend the identification models learned based on small-scale devices under LOS environments to large-scale devices under NLoS environments. Jian et al. \cite{jian2020deep} confirm that different environmental scenarios affect the identification accuracy, including channel conditions, SNR, number of devices, and training dataset size. Hence, the construction of RF fingerprints should consider multi-factors, including:
\begin{itemize}
\item \emph{NLOS Environments:} The authors of \cite{shen2022towards} and \cite{reus2020trust} consider the NLOS environments under indoor and outdoor scenarios, respectively.
\item \emph{Large-Scale Devices:} Liu et al. \cite{liu2020zero} use the USRP B210 to collect the ADS-B signals from 140 aircraft at Daytona Beach international airport. Shen et al. \cite{shen2022towards} use the USRP N210 to collect signals from 60 commercial off-the-shelf LoRa devices. Uzundurukan et al. \cite{uzundurukan2020database} employ a high sampling rate oscilloscope (Tektronix TDS7404) to record Bluetooth signals from 86 smartphones. Ya et al. \cite{ya2022large} utilize the Signal Hound SM200B to record 530 categories of long signals and 198 categories of short signals. Hanna et al. \cite{hanna2022wisig} use the USRP B210/N210/X310 to collect WiFi signals from 174 devices.
\item \emph{Multi-Receivers:} The authors of \cite{hanna2022wisig} and \cite{elmaghbub2021lora} employ 41 receivers, enabling Verification of collaborative authentication. 
\item \emph{Long Acquisition Time:} The acquisition time of \cite{al2020exposing}, \cite{hanna2022wisig}, and \cite{elmaghbub2021lora} is 10 days, 4 days, and 5 days, respectively.
\item \emph{Noisy Channel Environments:} Morin et al. \cite{morin2019transmitter} consider the dynamic channels interfered by mobile robots.
\item \emph{A Large Number of Fingerprint Samples:} The authors of \cite{sankhe2019oracle}, \cite{reus2020trust}, and \cite{elmaghbub2021lora} collect 2e7, 3e6, and 2e8 I/Q samples/individual, respectively.
\end{itemize}

Here, we provide the list of the open-source fingerprint datasets with the number and type of transmitters, waveform, type of receiver, and frequency in Tab. \ref{The Summarization of Open-Source RF Fingerprint Datasets}. 

\begin{table*}[h]
\caption{The Summarization of Open-Source RF Fingerprint Datasets}
\label{The Summarization of Open-Source RF Fingerprint Datasets}
\centering
\begin{tabular}{|A{0.6cm}|A{1.3cm}|A{2.3cm}|A{2.6cm}|A{2.5cm}|A{1.8cm}|}
\hline
\textbf{Ref.} & \textbf{Number of Transmitters} & \textbf{Type of Transmitters}  & \textbf{Waveform}      & \textbf{Type of Receiver}      & \textbf{Frequency}  \\
\hline
\cite{sankhe2019oracle}     & 16                              & USRP X310                      & IEEE 802.11a           & USRP B210                      & 2.45 GHz            \\
\hline
\cite{al2020exposing}     & 20                              & USRP X310/N210                 & IEEE 802.11a/g         & USRP N210                      & 2.432 GHz           \\
\hline
\cite{liu2020zero}     & 140                             & Aircraft                       & ADS-B                  & USRP B210                      & 1090 MHz            \\
\hline
\cite{shen2022towards}     & 60                              & Commercial LoRa devices        & LoRa                   & USRP N210                      & 868.1 MHz           \\
\hline
\cite{reus2020trust}     & 4                               & USRP X310                      & IEEE 802.11a/LTE/5G NR & USRP B210                      & 1.6 GHz $\sim$6 GHz \\
\hline
\cite{uzundurukan2020database}     & 86                              & Smartphones                    & Bluetooth              & Tektronix TDS7404              & 2.4 GHz             \\
\hline
\cite{ya2022large}     & 728                             & Aircraft                       & ADS-B                  & Signal Hound SM200B            & 1090 MHz            \\
\hline
\cite{hanna2022wisig}     & 174                             & WiFi transmitters              & IEEE 802.11a/g         & USRP B210/N210/X310            & 2462 MHz            \\
\hline
\cite{elmaghbub2021lora}     & 25                              & Pycom devices                  & LoRa                   & USRP B210                      & 915 MHz             \\
\hline
\cite{morin2019transmitter}     & 21                              & USRP N2932                     & IEEE 802.15.4          & USRP N2932                     & 400 MHz $\sim$4 GHz \\
\hline
\cite{martins2020drone}     & 17                              & Drone remote controllers (RCs) & Non-standard waveforms & Keysight MSOS604A oscilloscope & 2.4 GHz             \\
\hline
\cite{soltani2020rf}     & 7                               & DJI M100                       & Non-Standard           & USRP X310                      & 5 GHz   \\
\hline
\end{tabular}
\end{table*}

\subsection{Channel Fingerprint Datasets}
\label{Section V-B}
\subsubsection{Provided by Official Organizations}
\begin{itemize}
\item \emph{Industrial Datasets:} The National Institute of Standards and Technology (NIST) \cite{candell2017industrial} provide the CIR measurements collected in an outdoor facility with minimal interference from unexpected emitters and three indoor industrial facilities, including the automotive factory with large tracts of open areas, the steam generation plant with large machinery and overhead obstructions, and the machine shop with small space. The recorded CIRs are complex vectors of 8188 dimensions and contain rich channel information. The authors of \cite{xia2021multiple,pan2019threshold,meng2022multiuser,xie2021weighted,meng2022physical,liao2019multiuser,du2023physical} have utilized the dataset to verify the performance of proposed ML-based PLA schemes.
\item \emph{Generated based on real-world scenes from 40 Big Cities:} The China Academy of Information and Communications Technology (CAICT) provides the open-source dataset that can support a variety of wireless AI tasks \cite{chinamobile}, including sensing tasks such as location environmental reconstruction, MIMO tasks such as RIS and beam, and physical-layer tasks such as CSI feedback and channel estimation. The parameters can be customized to meet the needs of researchers.
\end{itemize}
\subsubsection{Provided by Individuals}
\begin{itemize}
\item \emph{4G LTE:} Jaeckel et al. \cite{jaeckel2014quadriga} extend the Wireless World Initiative for New Radio (WINNER) channel model to realize better trade off between complexity and accuracy. The proposed open-source model enables 3-D propagation, 3-D antenna patterns, scenario transitions, and variable terminal speeds. Gassner et al. \cite{gassner2021opencsi} present the first CSI-based radio map obtained by automated tools for LTE radio links.
\item \emph{MIMO:} To reproduce the stochastic properties of MIMO channels over time, frequency, and space, Liu et al. \cite{liu2012cost} provide the COST 2100 channel model, a stochastic channel model based on geometry. The COST 2100 channel is suitable for multi-user and distributed MIMO scenarios. 
\item \emph{Massive MIMO:} Alkhateeb et al. \cite{alkhateeb2019deepmimo} introduce the DeepMIMO dataset to advance the research of Massive MIMO and mmWave techniques. The dataset can be completely defined by the set of parameters.
\item \emph{WiFi:} Wang et al. \cite{wang2022framework} collect the CSI data in complex indoor environments at Colorado State University.
\item \emph{5G NR:} By considering the clustered delay line channel model \cite{3gpp2019study} that is suitable for link-level and system-level simulations, Zhang et al. \cite{zhang2021generalized} develop a generalized 5G NR dataset generator.
\item \emph{Reconfigurable Intelligent Surface (RIS):} Basar et al. \cite{basar2021indoor} consider the characteristics of the Reconfigurable Intelligent Surface (RIS), such as LOS probability, shadowing effects, shared clusters, and array responses, and further develop the SimRIS channel Simulator MATLAB package for the channel modeling of RIS-aided systems.
\item \emph{Outdoor Environments:} Alrabeiah et al. \cite{alrabeiah2020viwi} study the vision-aided wireless communications and provide the parametric, systematic, and scalable vision-wireless dataset framework. The high-fidelity synthetic wireless and visual data samples for the same scene can be generated through the advanced 3D modeling and ray tracing software.
\item \emph{Underwater Scenarios:} Qarabaqi et al. \cite{qarabaqi2013statistical} provide the underwater acoustic channel model, where the small-scale and large-scale effects are considered.
\end{itemize}

Here, we provide the list of the open-source channel datasets with the provider, environment, and descriptions in Tab. \ref{The Summarization of Open-Source Channel Fingerprint Datasets}.


\begin{longtable}{|A{0.7cm}|A{1.5cm}|A{2.1cm}|P{7.6cm}|}
\caption{The Summarization of Open-Source Channel Fingerprint Datasets} \label{The Summarization of Open-Source Channel Fingerprint Datasets} \\
\hline
\textbf{Ref.} & \textbf{Provider} & \textbf{Environment}& \textbf{Description} \\ \hline
\endfirsthead 
\hline
\textbf{Ref.} & \textbf{Provider} & \textbf{Environment} & \textbf{Description} \\
\hline
\endhead

\hline
\endfoot

\hline
\endlastfoot

\hline
\cite{candell2017industrial}     & NIST             & Industrial scenarios & The CIR measurements are collected under an outdoor environment and three typical industrial scenarios, including automotive factory, steam generation plant, and machine shop.              \\
\hline
\cite{chinamobile}    & CAICT            & City scenarios       & The dataset is generated based on real map scenes, including more than 1,000 scenes randomly selected from more than 40 big cities around the world.\\
\hline
\cite{jaeckel2014quadriga}     & \multirow{9}{*}[-30ex]{\begin{tabular}[c]{@{}A{1.5cm}@{}}Individual\end{tabular}}      & LTE                  & Provide a more realistic channel model that is extended from the popular WINNER channel model. The presented channel model can realize better trade-off between complexity and accuracy. \\
\cline{1-1} \cline{3-4}
\cite{gassner2021opencsi}     &                                       & LTE                  & Utilize USRP B200mini and a wheeled robot to record the CSI periodically.                                                                                                                                 \\
\cline{1-1} \cline{3-4}
\cite{liu2012cost}     & & MIMO                 & Propose a geometry-based stochastic channel model to reproduce the stochastic characteristics of MIMO channels over time, frequency, and space.\\
\cline{1-1} \cline{3-4}
\cite{alkhateeb2019deepmimo}     &                                       & Massive MIMO        & Present a generic Massive MIMO dataset based on ray-tracing data obtained from Remcom Wireless InSite \cite{chinamobile} for mmWave frequencies.\\
\cline{1-1} \cline{3-4}
\cite{wang2022framework}     &                                       & WiFi                 & provide a CSI dataset collected in complex indoor environments at Colorado State University.                                                                                                              \\
\cline{1-1} \cline{3-4}
\cite{zhang2021generalized}     &                                       & 5G NR                & Suggest a generalized channel dataset generator for 5G NR systems. The setting of different channel parameters and the generation of massive MIMO channels can be achieved.\\
\cline{1-1} \cline{3-4}
\cite{basar2021indoor}     &                                       & RIS, 5G              & Develop an accurate, open-source, and widely applicable RIS channel model for mmWave frequencies. Both indoor and outdoor environments are included. The 5G radio channel conditions are also considered.\\
\cline{1-1} \cline{3-4}
\cite{alrabeiah2020viwi}     &                                       & Outdoor              & The ViWi dataset framework is a parametric, systematic, and scalable data generation architecture. The high-fidelity synthetic wireless and vision data samples for the same scenes can be generated. \\
\cline{1-1} \cline{3-4}
\cite{qarabaqi2013statistical}     &                                       & Underwater           &Present an open-source underwater acoustic channel model that incorporates physical laws of acoustic propagation and the effects of inevitable random local displacements.\\    \hline       

\end{longtable}

\section{Challenges and Future Research Directions}
\label{Section VI}
\subsection{CVNN for CSI Fingerprints}
For MIMO and Massive MIMO systems, CSI fingerprints contain abundant spatial information of the channels between transmitters and receivers. A widely-used approach is to exploit CNNs to extract the inherent characteristics of CSI fingerprints: converting the CSIs into pictures with the dimensions of $N_R×N_T\times2$, where $N_R$ and $N_T$ respectively represent the number of antennas of the receiver and the transmitter and “2” indicates the real and imaginary parts. Hence, the complex CSIs are split into real and imaginary parts corresponding to different channels of CNNs, respectively. Considering that CNNs can not process complex data, they may not fully extract the complex features of MIMO channels, limiting the performance of the CSI-based PLA. Moreover, the future 6G systems are expected to support ultra-massive MIMO scenarios \cite{wang20206g}, where CSIs contain richer spatial features. Hence, how to effectively exploit CSI fingerprints to realize attack detection is an important issue.

To tackle this problem, one feasible solution is to exploit CVNNs, which can directly process complex signals. The higher computing complexity and more model parameters introduced by CVNNs can be addressed by network compression techniques.

\subsection{RIS-aided PLA}
Meng et al. \cite{meng2023efficient} propose a RIS-aided PLA architecture, where the configurable fingerprints can be obtained in NLoS environments. When the direct links between the transmitter and receiver are blocked, such design can enhance the robustness and reliability of fingerprints by creating alternative propagation paths. Meng et al. \cite{meng2023efficient} also verify that the parameters of RIS, such as number, magnitude, and phase response, influence the authentication performance. However, there are some issues to be addressed as follows: (1) How to derive the closed-form expression of false alarm rate and miss detection rate when the channel models are determined? (2) How to optimize the parameters of IRSs to obtain the optimal authentication performance?

\subsection{Game Between Legitimate Transmitters and Attackers}
Most state-of-the-art RL-based attack detection schemes primarily focus on the interaction between legitimate receivers and spoofing attackers, as demonstrated in \cite{xiao2016phy, li2015spoofing, zhou2022game}. However, effective defense strategies must also consider the role of trusted transmitters. Therefore, the game between legitimate transmitters and attackers represents a significant and valuable research direction.

For instance, envisioning an adversarial scenario involving multiple legitimate transmitters and spoofing attackers, a strategy emerges where legitimate transmitters nominate a representative to transmit authorized signals, aiming to minimize overall communication overhead. Simultaneously, attackers select representatives to transmit spoofing signals, attempting to evade detection by the legitimate receiver. The differing spatial locations of legitimate transmitters and attackers result in distinct location-specific channel fingerprints, influencing the similarity between them. This spatial discrepancy engenders a strategic game in the selection of representatives by both legal and illegal entities.

\subsection{Defense Against Multiple Cooperative Attackers}
Most of the state-of-the-art attack detection schemes study the defense against a single attacker, while a few schemes are designed against multi-attackers, such as \cite{xiao2016phy,yang2012detection,meng2023multidimensional}. The cooperation of multi-attackers is further studied in \cite{xiao2016phy}. Each spoofer can choose its attack frequency to maximize its utility, and multi-attackers cooperatively transmit spoofing signals to avoid collisions. The cooperation of multi-attackers is worth studying because it poses a more serious security threat to the communication system.
\subsection{Cross-Layer Authentication}
Although ML-based PLA can obtain high authentication performance, the PLA methods are not designed to replace the upper-layers authentication mechanisms. On the contrary, the PLA methods are provided to compensate for the upper-layers authentication mechanisms \cite{xie2020survey}. Hence, how to design a secure cross-layer authentication scheme or combine them effectively is an urgent problem. For example, Zhang et al. \cite{zhang2020fast} propose a lightweight cross-layer authentication for dynamic channel environments. The upper-layers authentication mechanisms help to update the parameters of the PLA model.
\subsection{ML for Key-based PLA}
The PLS technique includes key-based PLA, keyless PLA, and other security methods. Compared with the keyless schemes summarized in this paper, the key-based schemes utilize the channel reciprocity to generate keys between Alice and Bob. The channel reciprocity means that the same channel features can be observed at both ends of the same channel link. However, the uplink and downlink are in different frequency bands for frequency division duplexing (FDD) systems. Therefore, most of the key-based schemes designed for time division duplexing (TDD) systems are not suitable for FDD systems. On the other hand, FDD systems dominate existing cellular communications, such as narrowband IoT and 4G LTE. Hence, finding reciprocal channel characteristics in FDD systems is strongly desirable for the key-based PLA.

To address this issue, one promising method is to leverage ML techniques, especially DL techniques, to construct the reciprocal characteristics. For example, Zhang et al. \cite{zhang2021deep} propose a DL-based channel prediction framework to realize the estimation of features of one frequency band without any loopback.
\subsection{Interpretability of DL-Aided PLA}
Although DL-based PLA schemes can achieve high identification and detection accuracy, their interpretability is often inferior to that of traditional methods. Researchers frequently face difficulties in understanding the connections between the features extracted by DNNs and their prediction outcomes, a challenge commonly referred to as the "black-box problem" of DL technology. Therefore, it is crucial to develop DL-aided PLA schemes that are both secure and interpretable.

To solve the problem, Meng et al. \cite{meng2022multiuser} define the Fingerprint Library and provide the post-hoc explanations to answer the following question: which library examples explain the authentication results issued for a given fingerprint sample?

\subsection{Generative Large Model-Empowered PLA}
The generative large model excels in creating channel fingerprints and RF fingerprints due to its robust characterization and automatic feature extraction capabilities. By leveraging multi-level data abstraction, the model can effectively capture and depict the various nuances and fluctuations within wireless channels, resulting in the generation of fingerprint data with significant identification and utilitarian value. Nonetheless, it is vital to acknowledge the considerations surrounding the expenses associated with data acquisition, the intricate training demands of the model, and the necessity for robust fingerprints when implementing the technology in practical settings.


\section{Conclusion}
\label{Section VII}
This article has provided a comprehensive survey of the ML-based PLA in wireless communications. We have categorized the existing ML-based PLA schemes into two categories: multi-device identification and attack detection schemes. The former one, a multi-classification problem, aims to recognize which transmitter in the fingerprint database matches the received signal and is usually based on RF fingerprints. The latter one, a hypothesis testing problem, aims to identify the forged signal and is usually based on channel fingerprints. Moreover, we divided the DL-based multi-device identification schemes into several sub-categories: FCNN-based, CNN-based, RNN-based, Attention mechanism-based, data augmentation-based, CVNN-based, GAN-based, and AE-based schemes. We divide the ML-based attack detection schemes into three sub-categories: SL-based, UL-based, and RL-based schemes. We further summarized the open-source RF fingerprint and channel fingerprint datasets for researchers in related fields. At last, we concluded this paper with some recommendations and future research directions.

\section{Acknowledgements}
The work presented in this paper was supported by the National Key K\&D Program of China No. 2020YFB1806900, the National Natural Science Foundation of China No. 61871045, No. 62401074, No. 61932005, and No. U21A20448, Beijing Natural Science Foundation No. L242012 and No. L232051, the research foundation of Ministry of Education-China Mobile under Grant MCM20180101, and the Joint Research Fund for Beijing Natural Science Foundation and Haidian Original Innovation under Grant L232001.

\section{CRediT authorship contribution statement}
\textbf{Rui Meng:} investigation, methodology, and writing. 
\textbf{Bingxuan Xu:} investigation and writing (polish figures and tables).  
\textbf{Xiaodong Xu:} writing (review and editing), funding acquisition, and supervision. 
\textbf{Mengying Sun:} writing (review and editing). 
\textbf{Bizhu Wang:} writing (review and editing). 
\textbf{Shujun Han} writing (review and editing). 
\textbf{Suyu Lv} writing (review and editing). 
\textbf{Ping Zhang:} funding acquisition and supervision.















\section{Data availability}
No data was used for the research described in the article.

\bibliographystyle{elsarticle-num}
\bibliography{ref.bib}

\end{document}